\documentclass[twocolumn]{aastex701}

\usepackage{CJKutf8}
\usepackage [english]{babel}
\usepackage [autostyle, english = american]{csquotes}
\MakeOuterQuote{"}
\usepackage{etoolbox}
\usepackage{siunitx}
\usepackage{multirow}
\usepackage{booktabs}
\usepackage[capitalise,nameinlink]{cleveref}

\makeatletter
\AddToHook{cmd/section/before}{%
  \def\@currentcounter{section}%
  \cref@constructprefix{section}{\cref@result}%
}
\AddToHook{cmd/subsection/before}{%
  \def\@currentcounter{subsection}%
  \cref@constructprefix{subsection}{\cref@result}%
}
\AddToHook{cmd/subsubsection/before}{%
  \def\@currentcounter{subsubsection}%
  \cref@constructprefix{subsubsection}{\cref@result}%
}
\makeatother

\crefname{subsection}{Section}{Sections}
\Crefname{subsection}{Section}{Sections}
\crefname{subsubsection}{Section}{Sections}
\Crefname{subsubsection}{Section}{Sections}

\makeatletter

\patchcmd\H@refstepcounter{\protected@edef}{\protected@xdef}{}{}
\makeatother

\crefname{section}{Sec.}{Secs.}

\newcommand{\katachiprime}{Katachi-PRIME (\begin{CJK*}{UTF8}{min}形') \end{CJK*}}
\newcommand{\katachi}{Katachi (\begin{CJK*}{UTF8}{min}形) \end{CJK*}}

\submitjournal{ApJ}

\shorttitle{Is Morphology All You Need?}
\shortauthors{Alfonzo et al.}

\begin{document}


\title{Is Morphology All You Need?\\ Quantifying Imprints of Past Star Formation in the Present-Day Morphologies of Galaxies}

\author[0009-0007-3423-1332]{Juan Pablo Alfonzo}
\affiliation{Astronomical Institute, Tohoku University, 6-3, Aramaki, Aoba-ku, Sendai, Miyagi, 980-8578, Japan}
\email{juanpabloalfonzo@astr.tohoku.ac.jp}

\author[0000-0001-9298-3523]{Kartheik G. Iyer $\dag$}
\affiliation{Columbia Astrophysics Laboratory, Columbia University, 550 West 120th Street, New York, NY 10027, USA}
\affiliation{Center for Computational Astrophysics, Flatiron Institute, 162 Fifth Avenue, New York, NY 10010, USA}
\affiliation{Institute for Astronomy, University of Hawaii, 2680 Woodlawn Drive, Honolulu, HI 96822,USA}
\footnote{$\dag$ NASA Hubble Fellow}
\email{protopotato@gmail.com}

\author[0000-0002-2651-1701]{Masayuki Akiyama}
\affiliation{Astronomical Institute, Tohoku University, 6-3, Aramaki, Aoba-ku, Sendai, Miyagi, 980-8578, Japan}
\email{akiyama@astr.tohoku.ac.jp}

\author[0000-0003-4196-0617]{Camilla Pacifici}
\affiliation{Space Telescope Science Institute, 3700 San Martin Drive, Baltimore, MD 21218, USA}
\email{camilla.pacifici@gmail.com}


\author[0000-0002-5258-8761]{Abdurro'uf}
\affiliation{Department of Astronomy, Indiana University, Bloomington, IN 47405, USA}
\email{abdurroufastro@gmail.com}

\author[0000-0002-9217-1696]{Suchetha Cooray}
\affiliation{Kavli Institute for Particle Astrophysics and Cosmology, Stanford University, 452 Lomita Mall, Stanford, CA 94305, USA}
\affiliation{SLAC National Accelerator Laboratory, 2575 Sand Hill Road, Menlo Park, CA 94025, USA}
\affiliation{Astronomical Institute, Tohoku University, 6-3, Aramaki, Aoba-ku, Sendai, Miyagi, 980-8578, Japan}
\email{cooray@astr.tohokua.ac.jp}

\author[0009-0009-3404-5673]{Novan Saputra Haryana}
\affiliation{Astronomical Institute, Tohoku University, 6-3, Aramaki, Aoba-ku, Sendai, Miyagi, 980-8578, Japan}
\email{novan.haryana@astr.tohoku.ac.jp}

\author[0000-0002-1416-8483]{Marc Huertas-Company}
\affiliation{Instituto de Astrof\'isica de Canarias (IAC), La Laguna, E-38205, Spain}
\affiliation{Departamento de Astrof\'isica, Universidad de La Laguna (ULL), E-38200, La Laguna, Spain}
\affiliation{Observatoire de Paris, LERMA, PSL University, 61 avenue de l'Observatoire, F-75014 Paris, France}
\affiliation{Universit\'e Paris-Cit\'e, 5 Rue Thomas Mann, 75014 Paris, France}
\affiliation{Center for Computational Astrophysics, Flatiron Institute, 162 Fifth Avenue, New York, NY 10010, USA}
\email{mhuertas@iac.es}

\author[0000-0001-7964-5933]{Christopher C. Lovell}
\affiliation{Kavli Institute for Cosmology, University of Cambridge, Madingley Road, Cambridge CB3 0HA, UK}
\affiliation{Institute of Astronomy, University of Cambridge, Madingley Road, Cambridge CB3 0HA, UK}
\email{chris.lovell.astro@gmail.com}

\author[0000-0002-6748-6821]{Rachel Somerville}
\affiliation{Center for Computational Astrophysics, Flatiron Institute, 162 Fifth Avenue, New York, NY 10010, USA}
\email{rsomerville@flatironinstitute.org}

\author[0000-0003-2573-9832]{Joshua S. Speagle \begin{CJK*}{UTF8}{min}(沈佳士)\end{CJK*}}
\affiliation{Department of Statistical Sciences, University of Toronto, 9th Floor, Ontario Power Building, 700 University Avenue, Toronto, ON, M5G 1Z5, Canada}
\affiliation{Data Sciences Institute, University of Toronto, 17th Floor, Ontario Power Building, 700 University Avenue, Toronto, ON, M5G 1Z5, Canada}
\affiliation{David A. Dunlap Department of Astronomy $\&$ Astrophysics, University of Toronto, 50 St George Street, Toronto, ON, M5S 3H4, Canada}
\affiliation{Dunlap Institute for Astronomy $\&$ Astrophysics, University of Toronto, 50 St George Street, Toronto, ON, M5S 3H4, Canada}
\email{j.speagle@utoronto.ca}


\author[0000-0002-5077-881X]{John F. Wu}
\affiliation{Space Telescope Science Institute, 3700 San Martin Drive, Baltimore, MD 21218, USA}
\affiliation{Department of Physics $\&$ Astronomy, Johns Hopkins University, Baltimore, MD 21218,USA}
\email{jowu@stsci.edu}

\author[0000-0002-9851-2850]{Mikaeel Yunus}
\affiliation{Department of Physics $\&$ Astronomy, Johns Hopkins University, Baltimore, MD 21218,USA}
\email{myunus1@jh.edu}



\begin{abstract}

The next generation of galaxy surveys will bring a plethora of new observations, a vast majority of which will be observed with a few photometric broadbands, limiting the information we can infer from traditional SED fitting methods. To maximize the scientific value of these surveys, techniques that can go beyond traditional summary statistics in extracting information from galaxy observations are needed. Moving beyond integrated photometry, we introduce the \katachiprime method to do spectral energy distribution (SED) based inference that jointly incorporates morphological and SED information. \katachiprime uses implicit likelihood inference (ILI) to incorporate realistic morphological priors from \textit{gri} broadband images, to make predictions of: stellar mass (M$_*$), specific star formation rate (sSFR), redshift (z), age (t$_{50}$), dust attenuation (Av), and gas-phase metallicity (12+log(O/H)) of the galaxy. We find that this improves constraints compared to traditional 5-band photometry. We then use the learned morphological priors to examine the extent to which morphology encodes information about past star formation, finding a stark dichotomy in how morphology encodes star formation history. While the stable morphologies of main-sequence and quiescent galaxies can predict past star formation out to $\sim 4-7$ Gyr and $\sim3-5$ Gyr respectively, starburst and green-valley populations above and below the main sequence tend to experience structural disruptions that obscure all but the most recent star formation activity. Methods like \katachiprime maximize the scientific value of future telescopes by providing broad-baseline SED-level property estimates by incorporating morphological information.

\end{abstract}



\section{Introduction} \label{sec:intro}
Galaxy SEDs, a measurement of their flux density as a function of wavelength, are a sensitive probe of their present state and formation history. Composed of emission from their stars, ionized gas, dust, and active galactic nuclei (AGN), SEDs can be studied to infer the physical properties of galaxies, including their stellar mass, past star formation activity, dust content and metallicities, and are one of the most ubiquitous tools available to us to study the universe outside our own Galaxy.

SEDs usually come in one of two flavors: spectroscopic or photometric. Spectroscopic measurements provide highly detailed SEDs, which help to constrain physical parameter inference. However, collecting spectroscopic data is time-consuming and infeasible for most modern wide-area galaxy surveys. Photometric data give us a more coarse representation of a galaxy's SED, but can be collected in a fraction of the time, making them feasible for surveys with a large number of objects. Fitting observed data (both spectroscopic and photometric) to theoretical SEDs generated from stellar population synthesis models has been the gold standard for deriving the physical properties of galaxies \citep{FSPS,sed_fitting,conroy_sed,2019A&A...621A..51H,SED_Comp,iyer_sed}. 

However, there are some well-understood limitations to SED fitting. Two key ones being: the computational time SED fitting can take for a large number of galaxies, and parameter degeneracies that exist when inferring a galaxy's physical parameters when using a limited number of photometric bands. 

Traditional SED fitting methods, while robust, often involve comparing observed photometric or spectroscopic data to vast libraries of theoretical models, a task that can become prohibitively slow, especially with the advent of large-scale astronomical surveys \citep{sed_slow2,sed_slow}. This has lead to a rise in the use of amortized computation methods ranging from Brute force Bayesian computation using precomputed atlases \citep{cami_technique} to implicit likelihood inference (ILI), sometimes referred to as simulation based inference (SBI), to drastically reduce the computational time needed to do SED fitting at a catalog level \citep{ILI_speed_up2,ili_speed_up6,ILI_speed_up_3,ili_speed_up4,ltu-ili,ili_galaxy4,ili_speed_up5, ILI_sed_speed_up,2026MNRAS.547ag282H}. By learning the complex mapping between SEDs and physical parameters from forward models, the neural networks at the core of ILI create a highly efficient inference engine. This allows for the bypass of repetitive and time-consuming likelihood calculations or sampling steps typical of traditional Bayesian SED fitting for each new observation. By learning a direct mapping from observed SEDs to parameter posteriors, these methods enable rapid, yet principled, Bayesian inference for the enormous datasets that are defining modern astrophysics.

This, however, does not solve the other key limitation of SED fitting. The degeneracy problem in photometric, and to a lesser extent in spectroscopic, SED fitting. The degeneracy problem refers to the difficulty in uniquely determining the physical parameters of galaxies due to the presence of multiple combinations of these parameters that can produce similar SEDs. This issue arises from the inherent complexities in the relationship between a galaxy's physical properties (such as age, metallicity, star formation history, and dust attenuation) and the observed light across different wavelengths \citep{1997A&A...320...41K,dusty_starbust,2003PASJ...55..385T,2009ApJS..184..100L,2022ApJ...932...54N,2022MNRAS.511..765Q,2024MNRAS.531.3242M,2025MNRAS.tmp..792F,2025arXiv250213240S,2025ApJ...980..168D}. For example, a young galaxy with high dust content might produce a similar SED to an older galaxy. This overlap makes it challenging to pinpoint the exact characteristics of the galaxy. This is quite problematic since a majority of the data we collect of distant galaxies is photometric, as it is not feasible to collect spectra for a large number of objects. This means that the systematic uncertainties in our inferred physical properties might keep us from being able to fully understand the processes that shape galaxies over the course of their lives.  

Thankfully, SEDs are not the only observable we have of galaxies. Another important observable we have is high-resolution images of galaxies that capture their morphologies (i.e., their physical structure). An extensive body of literature, originating with Hubble's foundational classification \citep{1936rene.book.....H}, has categorized galaxies by morphological type, revealing a strong correlation with present-day star formation activity (\citealt{1998ApJ...495..139M, 2001AJ....122.1861S, 2005ApJ...629..143B, 2009ApJS..182..216K}; see reviews by \citealt{1994ARA&A..32..115R, 1998ARA&A..36..189K, 2014ARA&A..52..291C} and references therein). More recent data from observatories achieving higher redshifts and superior spatial resolution have further refined this relationship \citep{2008MNRAS.389.1179L, sfms_scatter, 2023ApJ...951..115E,2025arXiv250203532H}, and have also linked the morphological structure of galaxies to a suite of other properties such as their colors, environments, clustering, gas content, sizes, and kinematics \citep{1998ApJ...495..139M, 2005ApJ...629..143B, 2005ApJ...630....1Z, 2011MNRAS.413..813C, 2023MNRAS.521.1292P,2025ApJ...983...57W,2026arXiv260403503F}.

Despite our extensive understanding that galaxy morphology is intertwined with its physical properties, we traditionally do not use this information when inferring the physical properties of galaxies. Instead, morphological studies of galaxies have traditionally been seen as independent of SED fitting. The classic approach to galaxy morphology has been to use summary statistics to quantify morphology such as: S\'ersic index, concentration, asymmetry, smoothness, Gini, and M20 \citep{Morph_6, gini,concentration}. The pursuit of a single numerical summary for a galaxy's overall morphology means these approaches inevitably miss finer, yet significant, details. Critical features like bars, spiral arms, star-forming clumps, and the intricate structures within star-bursting or interacting systems are lost, thereby failing to account for the unique morphologies that often reflect a galaxy's evolutionary process. In doing this, we also destroy any unique information that could potentially help us to infer the physical properties of galaxies. This is reflected in the literature where simple summary statistics of the morphology have not strongly improved constraints on properties like redshift and stellar mass \citep{sed_morph,sed_morph3,sed_morph2}.

A more modern approach to this has been to deal with galaxy morphology through machine learning and, more recently, using deep learning \citep[see review by][]{Deep_Learning_Galaxy_Review}. Over the last decade or so, the literature has mostly tried to address this issue by using deep learning to categorize galaxies into discrete classes or categories \citep{Morph-class-3,Morph-class-4,Morph-class5,Morph-class-1,Morph-class-2}. These classes can be things like the presence of a prominent bulge, a disk-like structure, the presence of a bar, and so on. While this is more informative than classical summary statistics, it still suffers from the loss of information intrinsic to any sort of summary statistic.  

The established connection between morphology and a galaxy's physical state, coupled with the inability of summary statistics to capture its full complexity, has paved the way for innovative approaches. More recently, the literature reflects a surge in methods utilizing deep learning to directly interpret galactic morphology from high-resolution imaging to infer physical properties \citep{Wu_2019,2019A&A...622A.137B,buck_2021,Euclid_deep_learning_physical_prop,2024MNRAS.531.2011Z,katachi,Wu_2025,2025arXiv251112737A,2025arXiv251205078Y,Yunus2026}. These studies remarkably achieve accuracies on par with traditional photometric SED fitting and at a fraction of the computational time. This success strongly suggests that the visual morphology itself, with all its intricate details, offers a rich, largely untapped dataset for directly understanding the integrated physical parameters of galaxies. Furthermore, it indicates that the detailed visual appearance of galaxies is not just correlated with, but is a direct source of, quantifiable information about their fundamental physical parameters.

In this work, we introduce \katachiprime, a method that combines spectral and morphological information to help address the two key shortcomings of modern SED fitting. \katachiprime aims to build on \katachi \citep{katachi} by incorporating photometric data alongside morphological information in order to better constrain our predictions of galaxy physical parameters, and expand predictions to full posteriors in contrast to the median estimates from \citet{katachi}. It does so by leveraging two key machine learning methods: vision transformers and ILI. Vision transformers are used to get image embeddings, which are latent space vectors that can serve as a proxy for galaxy morphology. These embeddings are then used to train a neural posterior estimator to predict full posteriors of 6 key physical parameters (M$_*$, SFR, z, t$_{50}$, 12+log(O/H), Av). In doing so, \katachiprime has more constrained posteriors than traditional SED fitting using a comparable number of photometric bands. On top of this, it highlights the power of using galaxy morphology information in order to break degeneracies in the prediction of the physical parameters of galaxies. 

In this work, we explore two science cases to highlight the power of the method. Firstly, we explore how morphology acts as a "fossil record" of past star formation activity. We do so by tracing the "continuous memory" of the galaxies by examining how long the imprints of SFH last in the morphological structure of the galaxy. Secondly, we explore breaking the degeneracy of inferring physical properties of dusty starburst galaxies.

\katachiprime allows us to derive high-precision physical parameters of galaxies with limited photometric data in one specific part of the electromagnetic spectrum. This will be crucial for current and future large-scale surveys like Euclid, Rubin, and Roman, which will be largely photometric and limited to one part of the electromagnetic spectrum. Additionally, since \katachiprime is an ILI-based method, it can help to alleviate the high computational times that are needed to derive the physical parameters for the vast numbers of galaxies these surveys will target. 

This work aims to explore to what extent including the morphological information of galaxies can help constrain the inference of physical parameters. Furthermore, we explore if including morphological information can help alleviate degeneracy issues when inferring various physical parameters. In \cref{sec:data} we describe the data we used in this work as well as our sample selection. In \cref{sec:methods} we outline vision transformers and ILI with neural posterior estimators, which are the heart of the \katachiprime method. In \cref{sec:results} we show the results of the inference done by \katachiprime, and the unique abilities of the method compared to traditional SED fitting. In \cref{sec:discussion} we discuss these results and analyze what information can be extracted from the morphology of galaxies. 

Throughout this paper, we adopt a standard $\Lambda$ cold dark matter($\Lambda$CDM) cosmology with $\Omega_m = 0.3, \Omega_{\Lambda} = 0.7,$ and $H_0 =70$ km Mpc$^{-1}$ s$^{-1}$ .

\section{Data} \label{sec:data}

\subsection{GSWLC Survey} \label{sec:gswlc}
The GALEX-SDSS-WISE Legacy Catalog (GSWLC) is a crucial astronomical resource that merges data from three key surveys: Galaxy Evolution Explorer (GALEX), Sloan Digital Sky Survey (SDSS), and Wide-field Infrared Survey Explorer (WISE), to provide a comprehensive view of galaxy evolution \citep{gswlc}. It compiles physical properties for approximately 700,000 local universe galaxies, using ultraviolet, optical, and infrared observations. This multi-wavelength approach allows for the accurate determination of vital characteristics like star formation rates, stellar masses, and dust attenuation. 

The photometry for this dataset consists of 11 broadbands. FUV (135-174 nm) and NUV (175-280 nm) are provided from GALEX \citep{galex_filters}.\textit{u} (325-380 nm), \textit{g} (408-545 nm), \textit{r} (550-691 nm), \textit{i} (685-840 nm), and \textit{z} (865-960 nm) bands are provided from SDSS \citep{sdss_filters}. W1 (3.02-3.68 $\mu$m), W2 (4.08-5.12 $\mu$m), W3 (8.81-14.32 $\mu$m), and W4 (20.04-24.14 $\mu$m) are provided from WISE \citep{WISE_filters,WISE_filters2}. 

\begin{figure}
    \centering
    \includegraphics[width=1\linewidth]{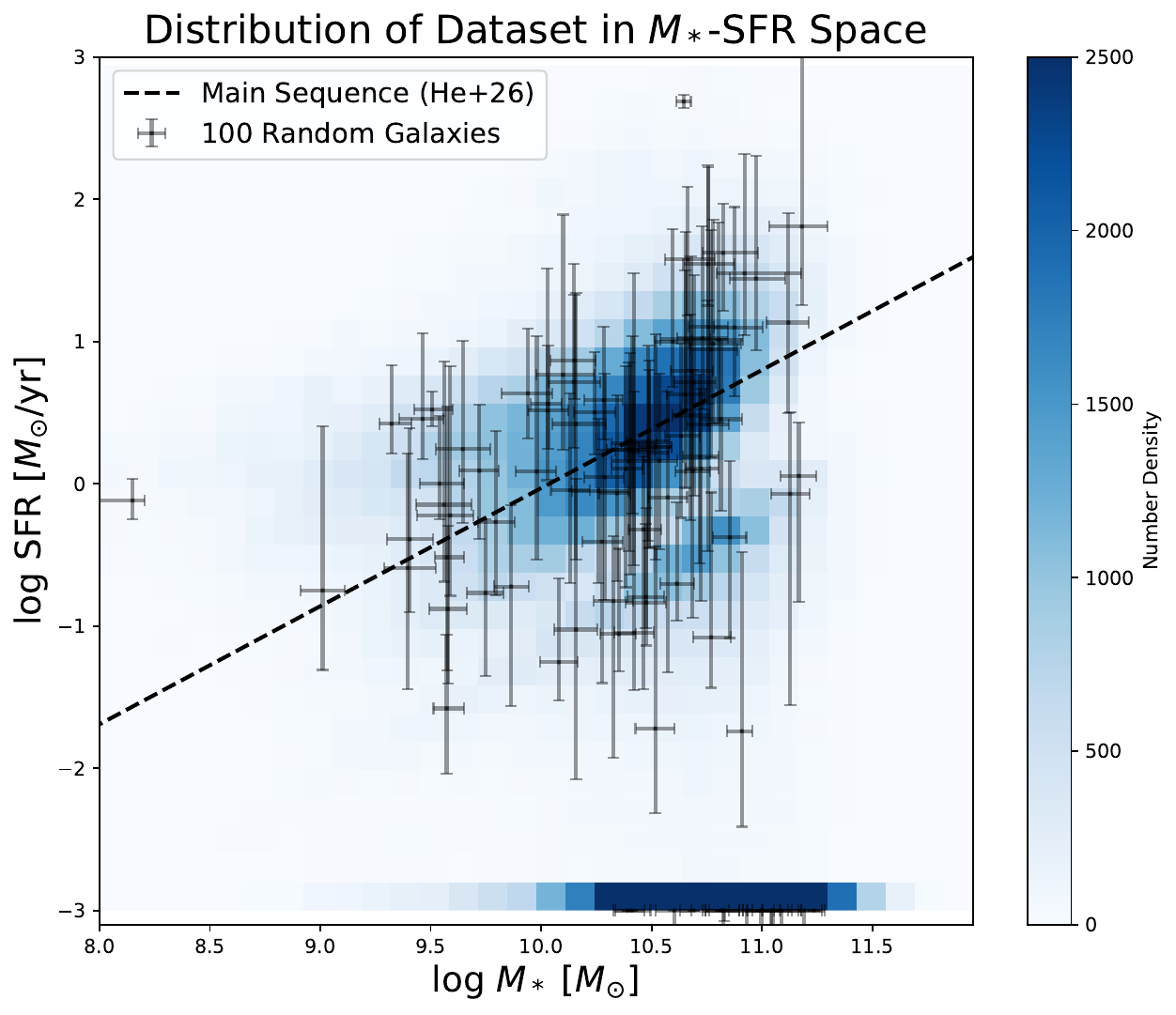}
    \caption{The number density of the galaxies in the sample used in this work (as described in \cref{sec:sed values}) on the M$_*$-SFR plane. 100 random galaxies from the sample are plotted with error bars showing the 16-84 confidence interval coming from the SED fitting. The main sequence from \cite{gswlc_sfms} is plotted as a dotted line.}
    \label{fig:num_density}
\end{figure}

\subsection{SED Fitting Values} \label{sec:sed values}

\subsubsection{Physical Parameters}
The sample in this work we follow the selection procedure of \cite{cami_sed}, which is limited to GSWLC galaxies with: 0.02 $<$ z $<$ 0.16, \textit{r} $<$ 22.2 mag, and that have GALEX and WISE W1 observations. This gave us a total sample of 232,343 galaxies. The number density of the galaxies in this sample on the M$_*$-SFR plane can be seen in \cref{fig:num_density}. 

To train the models in this work, we use physical parameters derived from traditional SED fitting as "true values". These values come from \cite{cami_sed} in which they fit 7  (FUV, NUV, g, r, i, z, and W1) of the 11-band photometry available for GSWLC galaxies to derive the galaxy's physical parameters\footnote{A schema of the derived values from \cite{cami_sed} can be found here: \href{https://github.com/camipacifici/local-gal-paper}{https://github.com/camipacifici/local-gal-paper}}. They omit fitting the u band from SDSS due to systematically fainter observed flux compared to what is predicted by the model. However, since GALEX NUV largely overlaps with the u band, the loss of information is minimal. W2, W3, and W4 bands are omitted from the fitting since ﬂuxes at longer wavelengths are contaminated by emission from warm dust, which is not included in their spectral modeling approach. 

They employ a comprehensive spectral modeling technique to derive physical parameters and star formation histories (SFHs) for a large sample of galaxies from their multi-band photometry. The approach is based on a method described in \cite{cami_technique}. The core process involves comparing the observed spectral energy distributions (SEDs) of galaxies to a large library of 1.5 million model SEDs. These model SEDs are built upon physically motivated SFHs and chemical enrichment histories generated from a post-treatment of the Millennium cosmological simulation. The models incorporate \cite{2003MNRAS.344.1000B} stellar population synthesis, consistent nebular emission calculated with CLOUDY \citep{cloudy}, and a detailed \cite{2000ApJ...539..718C} two-component dust attenuation prescription. 

A Bayesian approach is employed to calculate the likelihood of each model galaxy reproducing the observed SED of a given galaxy across 7 photometric bands. The spectral library was shown to accurately reproduce the distribution of observed galaxy colors in color–color space. The uncertainties on derived evolutionary stages (t$_{10}$, t$_{50}$, t$_{90}$) were quantified by fitting non-parametric SFHs from the semi-analytic model outlined in \cite{cami_SAM}. 

\subsubsection{Star Formation Histories} \label{sec:sfh_values}
The Dense Basis method \citep{dense_basis,dense_basis_2,dense_basis3} provides a non-parametric description of a galaxy's star formation history using its stellar mass, SFR, and a set of $t_X$ values that specify when a galaxy formed X\% of its total mass and provide constraints on the shape of the SFH. These values are used by a Gaussian process framework to create flexible SFHs that satisfy the integral constraints imposed by the $t_X$ values. While the method is generally used for spectral energy distribution (SED) modeling and fitting, it can also be used to reconstruct the SFHs of galaxies given their (M$_*$, SFR, $t_X$) tuples \citep{AGN_Feedback,katachi,sfms_scatter7}. We use M$_*$, SFR, t$_{90}$, t$_{50}$, and t$_{10}$ values from the SED fitting described in \cref{sec:sed values} to create SFHs for all galaxies in the sample using the Dense Basis method. In practice, because the fits developed in \citet{cami_sed} leave SFR in the last 10 Myr bin free (and decoupled from the rest of the SAM SFH), they can sometimes lead to unrealistically high sSFRs in the fits that are inconsistent with the rest of the SFH quantiles. For this reason, we include a regularization term when reconstructing the SFHs so that the (M$_*$, SFR, t$_{90}$, t$_{50}$, and t$_{10}$) tuple is self-consistent, and use these SFHs to train our models. 

In this work, we discretize the SFHs by extracting the SFR of the galaxies at the following lookback times (in Gyr): 0.1, 0.3, 0.5, 0.7, 0.8, 1.0, 1.3, 1.5, 1.7, 2.0, 2.5, 3.0, 3.5, 4.0, 4.5, 5.0, 5.5, 6.0, 6.5, 7.0, 7.5, 8.0, 8.5, 9.0, 9.5, and 10.0. These discretized versions of the SFHs are used for all results and analysis. 

\subsection{Morphological Parameters}
In this work, we use morphological summary statistics to complement our analysis. We use them to compare how much information they contain compared to full imaging data, as well as using them to show broad morphological population distributions in some of our figures.  We use traditional morphological summary statistics, such as S\'ersic index, from \cite{traditional_morph2} and \cite{traditional_morph}. In \cite{traditional_morph2} a catalogue of 2D, point spread function-corrected de Vaucouleurs, S\'ersic, de Vaucouleurs+Exponential, and S\'ersic+Exponential fits of $\approx$700,000 spectroscopically selected r-band galaxy images drawn from the SDSS Data Release 7 is provided. This analysis is extended to the g and i bands in \cite{traditional_morph}. In this work, we primarily use the S\'ersic index and half-light radius derived from the r band to complement our analysis. The results presented in this paper are unaffected by which band's S\'ersic index is used. 

We also use Galaxy Zoo-like labels for the galaxies in our sample from \cite{galaxy_zoo_labels}. They provide a morphological catalogue for $\approx$670 000 galaxies in the Sloan Digital Sky Survey, utilizing the Galaxy Zoo 2 \citep{galaxy_zoo_2} classification scheme alongside T-types, which provide a continuous numerical parameterization of the Hubble sequence. The classifications are obtained with Deep Learning algorithms using Convolutional Neural Networks (CNNs) that were trained on true labels from Galaxy Zoo 2.  In this work, we use the provided values of T-type and P$_{bulge}$ to complement our analysis. T-types range from -2 to 6 and correspond to the Hubble sequence, with smaller numbers being early-type galaxies and larger numbers being late-type galaxies. P$_{bulge}$ ranges from 0 to 1 and corresponds to how prominent or pronounced the bulge of a galaxy is. 

\subsection{SDSS Imaging}
We obtain 64 x 64 pixel cutouts of the GSWLC sample from \cref{sec:gswlc}, using the SDSS \textit{gri} broadband images of these galaxies with a pixel scale of 0.262 arcseconds per pixel. This means the cutouts in our sample span a physical width of approximately $6.8\text{ kpc}$ to $46.2\text{ kpc}$ (with a median of $\approx 25.3\text{ kpc}$ at $z = 0.08$). These images were converted to RGB using the pipeline described in \cite{RGB_Images}, which preserves relative flux. These images were acquired using the SDSS SkyServer via the Legacy Survey website\footnote{\href{https://www.legacysurvey.org/}{https://www.legacysurvey.org/}}. We use images from SDSS DR8 \citep{dr8}.

\section{Methods} \label{sec:methods}
\subsection{Vision Transformers}
Vision transformers (ViTs) are a type of neural network architecture that have recently emerged as powerful models for computer vision tasks. They are based on the transformer architecture, which was originally developed for natural language processing (NLP) tasks \citep{attention_is_all_you_need}.  

ViTs process images by dividing them into a sequence of patches. Each patch is then flattened into a vector. These flattened patch vectors are then linearly projected into a lower-dimensional embedding space. Since the Transformer architecture itself does not inherently understand the spatial order of the patches, positional encodings are added to the patch embeddings. These encodings provide the model with information about the original position of each patch within the image. This allows the model to learn a wide range of high-level features and correlate them using attention, something that can be more difficult to do with standard convolutional neural network-based approaches.   

The sequence of embedded patches (including the positional encodings) is then fed into a standard Transformer encoder. The transformer encoder is a stack of self-attention layers (referred to as "Multi-head"). This is the core of the Transformer. The self-attention mechanism allows each patch embedding to interact with and weigh the importance of all other patch embeddings in the sequence. This enables the model to capture long-range dependencies and global context within the image. "Multi-head" means that this attention process is performed multiple times in parallel with different learned linear projections, allowing the model to focus on different aspects of the relationships between patches. This is important for image classification and regression, as it allows the model to learn how the different parts of an image contribute to its overall prediction target.

ViTs have shown state-of-the-art performance on a variety of image classification benchmarks \citep{vitperformance,vitperformance3,vitperformance4,vitperformance2}. ViTs can be more efficient when global context is crucial and when trained on large datasets than convolutional neural networks (CNNs), which are the traditional models for image classification \citep{vitvcnn,vitvcnn3,vitvcnn2}. ViTs have been shown to be effective at galaxy inference tasks such as predicting physical properties or morphological classes \citep{vit_galaxies,astroclip,aion1,astropt,galaxy_scailing_laws}.

In this work, we used a pre-trained ViT on ImageNet-21k (14 million images, 21,843 classes) at resolution 224x224, and fine-tuned on ImageNet 2012 (1 million images, 1,000 classes) at resolution 224x224 \citep{pretrained_vit}. We fine-tuned the model on galaxy images by having the ViT with a regression head predict the M$_*$ and sSFR of the galaxies in the training sample. We found that the image embeddings from the fine-tuned ViT yield significantly better predictions ($\geq 0.1$ dex improvement for predicted parameters) of galaxy properties than using a non-fine-tuned ViT. This fine-tuning step allows the ViT to understand unique patterns found in galaxy images that are associated with the galaxies' morphology. 

Therefore, we use the image embeddings from a pre-trained and fine-tuned ViT in this work. We use the Transformers library \citep{transformers} through the HuggingFace API to obtain our image embeddings.   

\subsection{Implicit Likelihood Inference} \label{sec:ILI}
Implicit likelihood inference (ILI) enables Bayesian inference when analytical likelihoods are intractable by directly approximating the conditional posterior distribution. In this work, we implement ILI via neural posterior estimation (NPE) using normalizing flows \citep{normalizing_flow,normalizing_flow2} to learn the posterior $q_{\phi}(\theta \mid x)$, where $x$ represents the observational data (photometry and images) and $\theta$ denotes the target physical parameters ($\text{M}_*$, $\text{sSFR}$, $z$, $t_{50}$, $A_v$, and $12+\log(\text{O/H})$) as well as the SFH of the galaxy. The network is trained by minimizing the negative log-probability on a dataset of $(\theta_i, x_i)$ pairs derived from SED fitting as described in \cref{sec:sed values}, assuming uniform priors. Once optimized, the network functions as an amortized estimator. This allows for the rapid generation of posterior distributions for new observations ($x_{obs}$) and bypasses computationally expensive traditional sampling methods like MCMC, efficiently yielding parameter samples to estimate means, credible intervals, and underlying correlations.
 
 For the interested reader, we suggest \cite{NPE1} and \cite{NPE2} for a more in-depth and technical summary of ILI using NPEs, and \cite{ILI_speed_up2,2023ApJS..268....7W, ltu-ili} for a more technical description of ILI in an astrophysics context. 

In this work, we take 80$\%$ of our complete sample (N=185,874) to serve as the training data for the neural network and reserve the remaining 20$\%$ (N=46,469) as a test data set to evaluate the performance of the trained models. The train and test data sets are mutually exclusive, meaning that the test set has galaxies that the trained ILI has never seen before. 

We also experimented with other training set sizes to see how it affected model performance. Even when only using a training split of 10$\%$ (N=23,234), we find comparable accuracy on the test set as when we used the 80$\%$ training split. 

We use an ensemble of normalizing flows consisting of a MAF (Masked Autoregressive Flow) and a NSF (Neural Spline Flow), each with 128 hidden features and 10 transforms. We use a batch size of 64 and a learning rate of $10^{-4}$.  We use the LtU-ILI library \citep{ltu-ili} to train all ILI models in this work. 

\subsection{UMAP}

To visualize and explore the underlying structure of the high-dimensional image embeddings, we employed the Uniform Manifold Approximation and Projection (UMAP) algorithm \citep{UMAP}. UMAP is a non-linear dimensionality reduction technique that constructs a low-dimensional representation of the data while preserving as much of its high-dimensional topological structure as possible, particularly local connectivity.

For this study, UMAP was applied with n$\_$neighbors = 15 and min$\_$dist = 0.1, using the Euclidean distance metric. The resulting low-dimensional embedding allows for qualitative assessment of what morphological features the ViT learns, as well as their connections to physical parameters. The full results of this analysis can be seen in \cref{sec:umap_results}. 

\subsection{Pipeline for \katachiprime: Predicting Galaxy Properties from Photometry and Images} \label{sec:pipeline}
The generalized pipeline introduced and used in this work for predicting galaxy properties from their \textit{gri} broadband images is as follows: 

\begin{enumerate}
    \item Select a subsample of galaxies that is at least 10,000 galaxies. Ensure that these galaxies are a representative sample of the complete sample (i.e., make sure there is no significant bias in the distribution of physical parameters in the subsample vs. the full sample). Preferably, this should be a "golden sample" with broad wavelength coverage, spectra, or both. This will be the "training sample" which will be used for training the ILI. 
    \item Generate "ground truth" that will be used for training using a SED fitting code of choice. Fit the photometry or spectra of the galaxies to get all the physical parameters you are interested in predicting. 
    \item Fine-tune the pre-trained ViT by training it to predict a pair of the physical parameters (in this work M$_*$ and SFR) derived in step 2.  
    \item Run the galaxy images through the ViT to generate image embeddings. The image embeddings are 768-dimensional vectors that summarize the patterns and structures in the image. In this case, we can treat them as a proxy for the morphology of the galaxy. 
    \item Train the ILI to predict the SED fitting derived physical parameters using the galaxy's photometry and image embeddings as inputs\footnote{Note that for some cases in this paper we only train using the photometry OR only the image embeddings. This is done to see how the data perform independently of each other. For best results, using both in conjunction is advised. This is discussed in detail in \cref{sec:results} and \cref{sec:discussion}.}. All the data for training should come exclusively from the training sample.
    \item Validate the inferred posteriors generated by the model by running Tests of Accuracy with Random Points (TARP). TARP estimates the coverage probabilities of generative posterior estimators \citep{TARP}. 
    \item After the ILI has been trained and validated, pass it the photometry and image embeddings of the rest of the sample to infer posteriors for the physical parameters
\end{enumerate}

\begin{figure*}
    \centering
    \includegraphics[width=0.9\linewidth]{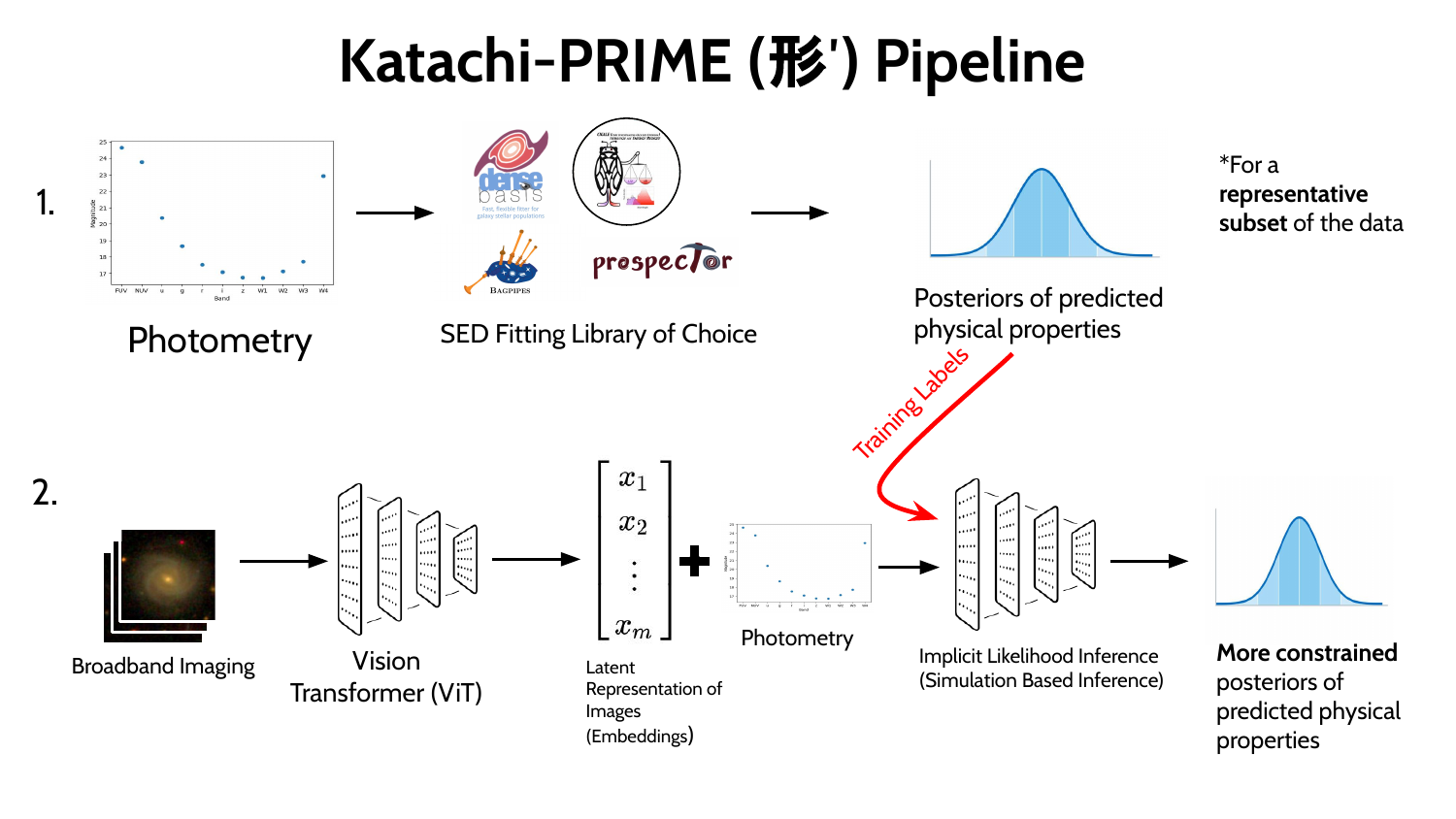}
    \caption{An overview of the \katachiprime method. \katachiprime is able to infer physical properties of galaxies with tighter posteriors than traditional SED fitting by leveraging the information contained in galaxy morphology. Running the galaxy images through a vision transformer allows us to track patterns in the morphology of galaxies. These vectors, which encode the morphology of the galaxies, are combined with available photometry and used to do implicit likelihood inference for physical parameters (M$_*$, SFR, t$_{50}$, 12+log(O/H), and Av) of galaxies.}
    \label{fig:pipeline_diagram}
\end{figure*}

A visual summary of this process can be seen in \cref{fig:pipeline_diagram}. The name of this method is \katachiprime as it builds upon the idea of the \katachi method presented in \cite{katachi}. The code for this pipeline can also be found in the GitHub repository associated with this project\footnote{\href{https://github.com/juanpabloalfonzo/Katachi-PRIME}{https://github.com/juanpabloalfonzo/Katachi-PRIME}}. Additionally, all data necessary to recreate every result in this paper can be found on HuggingFace\footnote{\href{https://huggingface.co/juanpabloalfonzo}{https://huggingface.co/juanpabloalfonzo}}. The code in the GitHub repository is designed to automatically access the data from HuggingFace to run the analysis shown in this paper. 

The training time for fine-tuning the ViT in this work was around 40 minutes using a single NVIDIA RTX A6000 GPU. The training time for each of these ILI models ranges from 30 to 33 hours each on a single NVIDIA RTX A6000 GPU, depending on the input type used. 

For a comparison of \katachiprime and \katachi on this dataset, refer to Appendix \ref{app:katachi}. 

\section{Results} \label{sec:results}

For this work, we primarily focus on four independently trained ILI models in order to compare how the different input data types affect physical parameter inference. The four models vary by the type of data that was used as an input. All four models were trained to predict the same set of physical parameters (M$_*$, sSFR, z, t$_{50}$, Av, 12+log(O/H)) and the discretized SFHs derived from SED fitting as described in \cref{sec:sed values}. The four types of inputs used are as follows: 

\begin{itemize}
    \item \textit{gri} Broadband Imaging ("Image Model")
    \item 5-band Optical Photometry (u, g, r, i, z)  
    \item \textit{gri} Broadband Image + 5-band Optical Photometry ("Combined Model")
    \item 11-band UV-Optical-IR Photometry (FUV, NUV, u, g, r, i, z, W1, W2, W3, W4)
\end{itemize}

The following results section discusses mainly the first three models listed, while some other ones are introduced in the discussion section to further evaluate the information contained in the different input modalities. 

We also sometimes refer to a combined model that uses imaging data and 11-band photometry data as an input. The discussion of this model along with the 11-band photometry model can be found in Appendix \ref{app:accuracy} and Appendix \ref{app:corner}. 

In \cref{sec:sfh} we discuss a model that takes \textit{gri} images as inputs, but which predicts the SFHs of the galaxies alongside the aforementioned set of 6 physical parameters. The true values of SFHs used to train this model come from the Dense Basis method as described in \cref{sec:sfh_values}. This model is referred to as the "SFH model". 

For all models in this work, we run TARP to validate our inferred posteriors and ensure they have statistical convergence. 

All results shown in this paper come from the full dataset (both the training and test set combined). We verify that the trends reported here are the same even if we look at just the test set in isolation. We chose to look at results across the full dataset in order to get the most robust statistical results by having the largest sample size possible. This approach allows us to check that the model is not overfitting, while also allowing our results to be more statistically significant. 

\subsection{Can Morphology Predict Galaxy Properties Accurately?}

\subsubsection{Inferred Posteriors}
\begin{figure*}
    \centering
    \includegraphics[width=1\linewidth,trim = 5cm 0cm 5cm 0cm,clip]{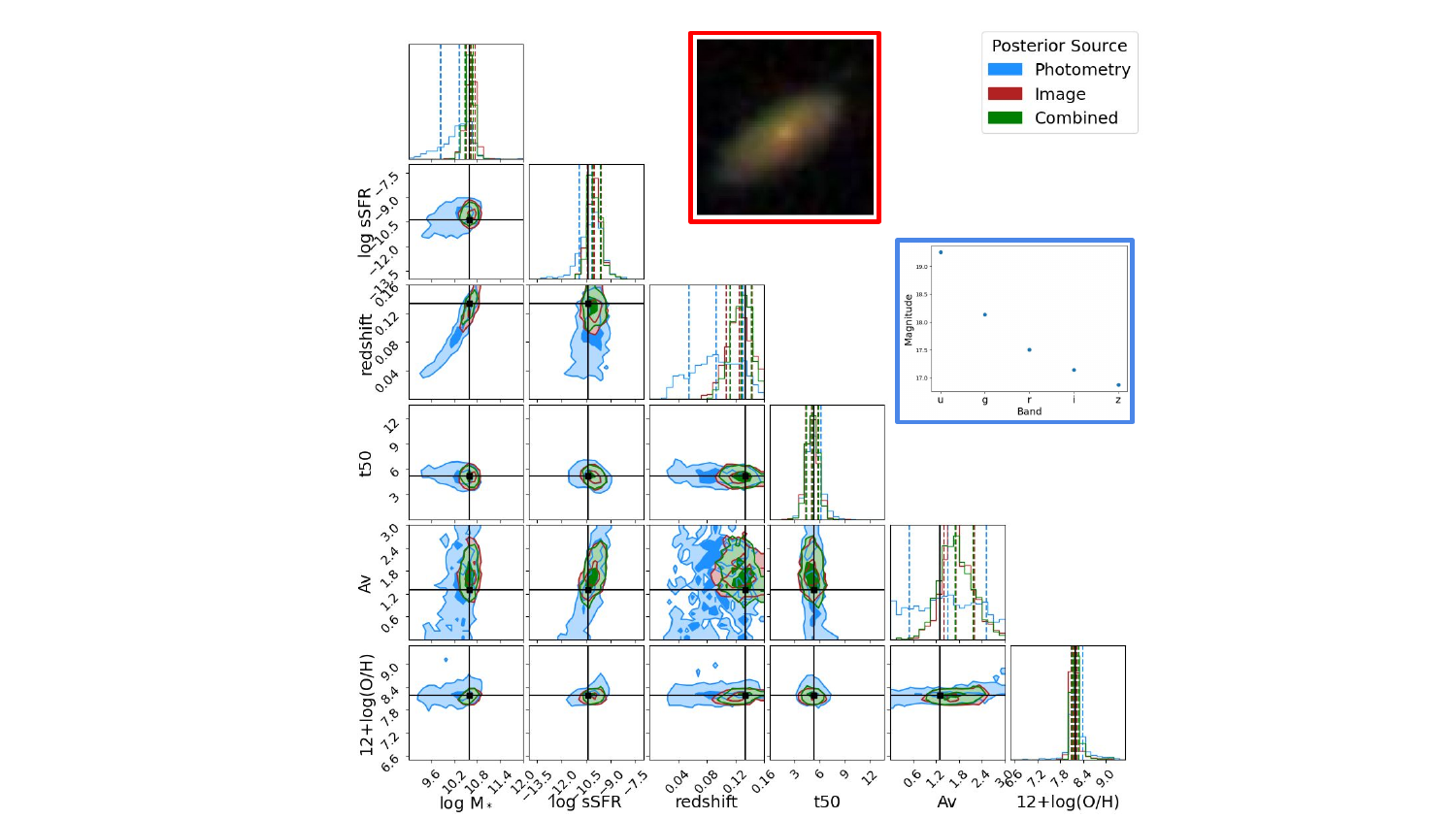}
    \caption{The posteriors inferred by the image model (red), 5-band photometry model (blue), and combined model (green) for all 6 predicted physical parameters of a randomly selected galaxy from the test sample. The image model and photometry model show comparable accuracy, but the image model shows more constrained posteriors and joint probabilities. The combined model shows the best accuracy and most constrained confidence intervals. This suggests that morphology and photometry hold some mutually exclusive information that can be combined to better constrain the inference of the physical properties of galaxies.}
    \label{fig:single_galaxy_image_and_photometry_posterior}
\end{figure*}

To illustrate the full capability of the \katachiprime method, we randomly select a galaxy from the test set to examine the inferred posterior. We focus on how different input modalities affect the inferred posteriors by comparing the posteriors of the different ILI models. To compare the posteriors for the 6 physical parameters of the image model, the 5-band photometry model, and the combined model, we create a corner plot with their predictions. This can be seen in \cref{fig:single_galaxy_image_and_photometry_posterior}. The posteriors are colored depending on which model (or which input types) generated them. Red indicates predictions made using exclusively 3-band imaging data, blue indicates predictions made using exclusively 5-band photometry data, and green indicates predictions made using both 3-band imaging and 5-band photometric data.  

We can see in \cref{fig:single_galaxy_image_and_photometry_posterior} that although the accuracy of the posteriors of the imaging and 5-band photometry models are quite similar, the image model provides slightly tighter posteriors. This trend is also verified across the test set to ensure this galaxy is not a unique case. As is further discussed in \cref{sec:pred_accuracy}, it seems that the information contained in galaxy morphology is enough to compensate for the information missing from having 2 fewer photometric bands. Furthermore, we see that combining both data sources yields the best accuracy as well as the tightest posterior distributions. This suggests that morphology and photometry hold mutually exclusive information that can be combined to better constrain the inference of the physical properties of galaxies.

A similar plot for the 11-band photometry model can be seen in Appendix \ref{app:corner}. Unlike the 5-band photometry model, the 11-band photometry model shows far better precision than the image model. Additionally, looking at the combined imaging and 11-band photometry model, imaging data does not help constrain predictions as much for 11-band photometry as it does for 5-band photometry. This is likely due to the fact that the additional information morphology provides to 5-band photometry is less than is contained in the additional 6 photometric bands in the 11-band photometry, which span a wide range of the electromagnetic (EM) spectrum. 

\subsubsection{Prediction Accuracy and Precision} \label{sec:pred_accuracy}

In this section, we explore how well the ILI models are at predicting the 6 different physical parameters of galaxies on a population level. The accuracy in prediction is calculated by assessing the bias ($\mu$). The precision of the prediction is calculated by assessing the scatter ($\sigma_{\text{NMAD}}$) of each model. The residuals for each galaxy are computed by sampling the posterior 1000 times, then calculating the difference between the median of these 1000 samples and the true value for each of the inferred parameters. The bias is the median value of these residuals across the full data set. The scatter is calculated by taking the absolute value of the difference between the bias and the residuals. In this work, we use Normalized Median Absolute Deviation (NMAD) as our measure of scatter. We use NMAD rather than standard deviation because NMAD is less sensitive to outliers. The bias highlights whether any systematic offsets are present in the models, while the scatter highlights the overall precision of the predictions. The values derived from the 7-band SED fitting are taken as the true value for the calculation of the bias and scatter (as described in \cref{sec:sed values}). 

\begin{figure*}
    \centering
    \includegraphics[width=0.85\linewidth]{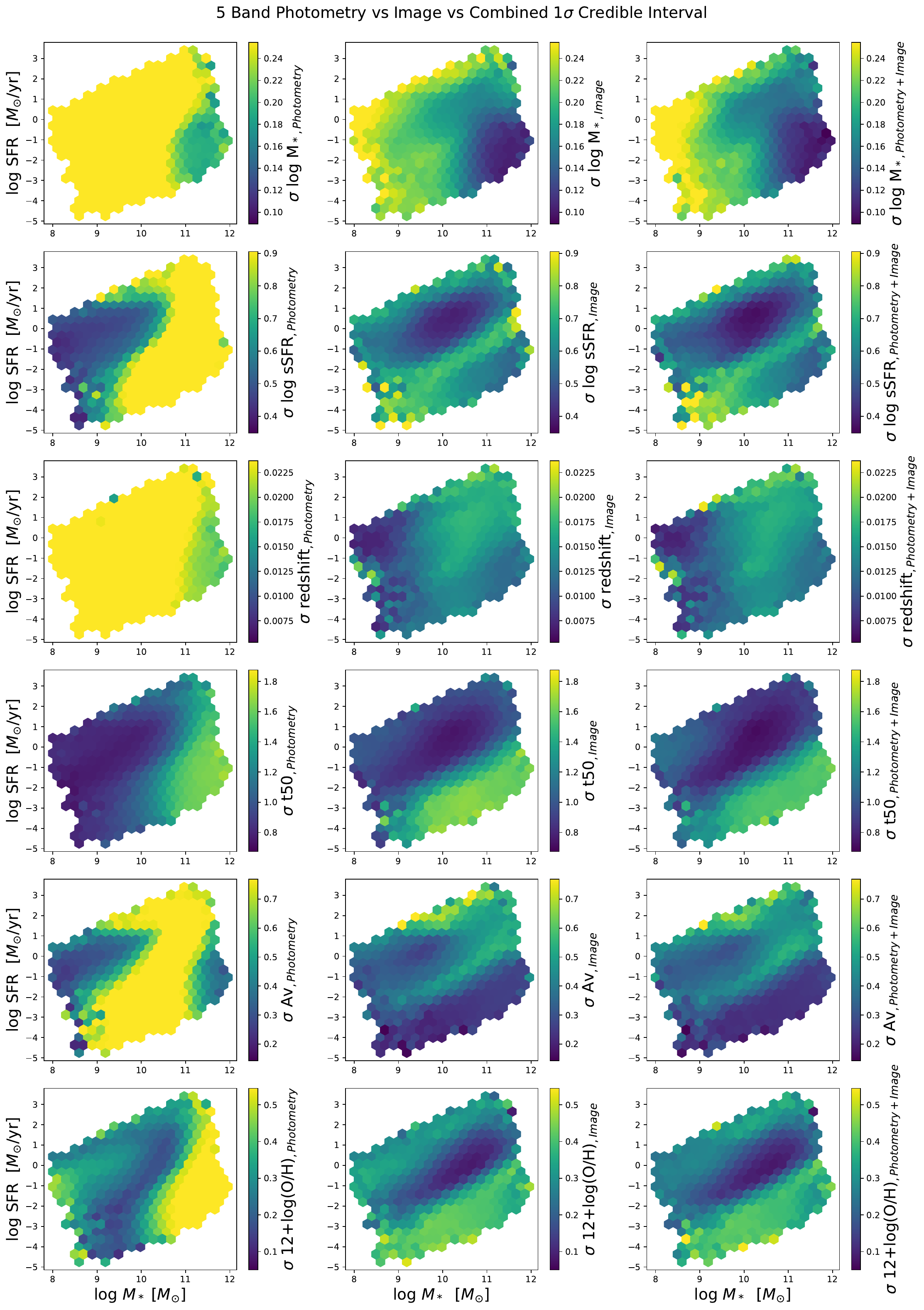}
    \caption{The 1$\sigma$ credible interval of the inferred posterior of the 5-band photometric model vs the image model vs the combined model in relation to the galaxy's location on the M$_*$-SFR plane. The first column shows the precision for the 5-band photometric model, the second column shows the precision for the image model, and the third column shows the precision of the combined model. Each row corresponds to one of the 6 inferred parameters (M$_*$, sSFR, z, t$_{50}$, Av, 12+log(O/H)). The $1\sigma$ errors in the left column are highly dependent on the galaxy's location on the $M_*$-SFR plane, whereas the image and combined model in the right columns provide much more stable, uniform precision regardless of the location on the M$_*$-SFR plane.}
    \label{fig:5_band_accuracy}
\end{figure*}

A comparison of how precise the image model is compared to the 5-band photometry model, the combined model, and the 11-band model on a population level can be seen in \cref{tab:accuracy}. In \cref{fig:5_band_accuracy}, the galaxies are plotted on the M$_*$-SFR plane and are colored by the 1$\sigma$ credible interval of the inferred posterior for each of their physical parameters. Each of the 6 rows in the plot represents a different physical parameter, with the first column showing results for the 5-band photometry model, the second column showing results for the image model, and the third column showing the results for the combined model.

\cref{fig:5_band_accuracy} illustrates the prediction precision across the M$_*$-SFR plane, highlighting the fundamental degeneracy limits of the 5-band photometry model (leftmost column). Notably, the $1\sigma$ credible intervals for log sSFR broaden significantly for galaxies transitioning off the main sequence into the quiescent population. Similarly, the precision of dust attenuation (Av) degrades severely for high-mass star-forming galaxies, reflecting the inability of integrated broadband SEDs to break the classical age-dust degeneracy in these regimes. In contrast, the imaging and combined models (rightmost columns) do not have these regions of high uncertainty. By incorporating resolved spatial morphology, the network successfully breaks these physical degeneracies, yielding tight, uniform constraints for both quiescent and massive obscured galaxies across the entire parameter space.

A comparison of the precision between the image model and the combined model can be seen in the second and third columns of \cref{fig:5_band_accuracy}. Across all the inferred parameters, we see marginal improvement in the performance of the combined model when compared to the image model. They essentially struggle with similar populations of galaxies across the M$_*$-SFR plane. However, looking at \cref{tab:accuracy}, we see that the combined model does outperform the image model on a population level.

Overall, we see a significant improvement in prediction scatter from the imaging and combined model over the 5-band photometric model, and only a marginal improvement from the combined model over the image model. This suggests that the major improvement from the combined model is being driven by the additional imaging data. 

This is to say the inclusion of the galaxy’s morphology provides complementary and unique information that is not found in the two additional optical photometric bands. Furthermore, these results suggest that the information contained in the galaxy's morphology is greater than the information that is contained in these two additional bands when it comes to inferring these physical parameters. 

\begin{table}[h!]
    \centering
    \resizebox{\columnwidth}{!}{
    \begin{tabular}{llcc}
    \toprule
    \textbf{Model} & \textbf{Parameter} & \textbf{$\mu$} & \textbf{$\sigma_{\text{NMAD}}$} \\
    \midrule
    
    \multirow{6}{*}{Imaging} 
    & $\log M_*$            & $\mathbf{0.000}$& $0.154$\\
    & $\log \text{sSFR}$     & $0.028$& $0.438$\\
    & $z$                   & $\mathbf{-0.001}$& $0.014$\\
    & $t_{50}$              & $-0.030$& $0.931$\\
    & $A_V$                 & $\mathbf{-0.004}$& $0.303$ \\
    & $12+\log(\text{O/H})$ & $\mathbf{0.000}$ & $0.124$\\
    
    \midrule
    \multirow{6}{*}{5-band Photometry} 
    & $\log M_*$            & $-0.036$  & $0.257$ \\
    & $\log \text{sSFR}$     & $\mathbf{0.020}$ & $0.741$ \\
    & $z$                   & $-0.004$ & $0.025$ \\
    & $t_{50}$              & $-0.046$  & $0.940$ \\
    & $A_V$                 & $0.044$ & $0.617$ \\
    & $12+\log(\text{O/H})$ & $0.025$  & $0.174$ \\
    
    \midrule
    \multirow{6}{*}{Combined} 
    & $\log M_*$            & $0.009$  & $\mathbf{0.145}$ \\
    & $\log \text{sSFR}$     & $-0.024$ & $\mathbf{0.395}$ \\
    & $z$                   & $\mathbf{0.001}$  & $\mathbf{0.013}$ \\
    & $t_{50}$              & $\mathbf{-0.029}$ & $\mathbf{0.897}$ \\
    & $A_V$                 & $-0.026$ & $0.296$ \\
    & $12+\log(\text{O/H})$ & $0.007$  & $0.113$ \\

    \midrule
    \multirow{6}{*}{11-Band Photometry} 
    & $\log M_*$            & $-0.019$  & $0.160$ \\
    & $\log \text{sSFR}$     & $0.028$ & $\mathbf{0.361}$ \\
    & $z$                   & $0.002$  & $0.017$ \\
    & $t_{50}$              & $\mathbf{-0.006}$ & $0.909$ \\
    & $A_V$                 & $0.032$ & $\mathbf{0.170}$ \\
    & $12+\log(\text{O/H})$ & $0.006$  & $\mathbf{0.096}$ \\
    
    \bottomrule
    \end{tabular}
    }
     \caption{A comparison of the bias and scatter (as described at the start of \cref{sec:pred_accuracy}) for the 4 main trained ILI models described in this work. The column titles are the type of input data the models use. All models show no systematic errors (near 0 bias values). The combined model achieves better accuracy (lowest scatter) for all parameters compared to the 5-band photometry and image model. The image model and 11-band model show comparable performance.}
    \label{tab:accuracy}
\end{table}

For a similar plot to \cref{fig:5_band_accuracy} that compares the accuracy of the 11-band photometry model as well as a combined imaging and 11-band photometry model, refer to Appendix \ref{app:accuracy}.    

\subsection{Morphology Helps Break Age-Dust-Metallicity Degeneracies} \label{sec:degeneracy_results}
In the previous sections, we discussed how the \katachiprime method uses morphology to better constrain the predictions of the physical parameters of galaxies. In a similar vein, the method is able to show that morphological information can help break well-known degeneracies between predicted physical parameters. 

In order to systematically explore the degeneracy that exists when predicting each pair of physical parameters across our sample, we compute the covariance of the joint probabilities. We quantify the degeneracy between predicted physical parameters by computing the Spearman rank correlation coefficient ($\rho$) defined as: 

\begin{equation}
     \rho = 1 - \frac{6 \sum_{i=1}^{n} d_i^2}{n(n^2 - 1)}
\end{equation}

where $d_i$ is the difference between the ranks of the i-th corresponding values of the two parameters, and $n$ is the number of pairs of observations. $\rho$ can take values from -1 to 1. -1 denotes a perfect anti-correlation between the two parameters, and 1 denotes a perfect correlation between the two parameters. We compute $\rho$ for each parameter pair for every galaxy in our data set. 

\begin{figure}
    \centering
    \includegraphics[width=1\linewidth]{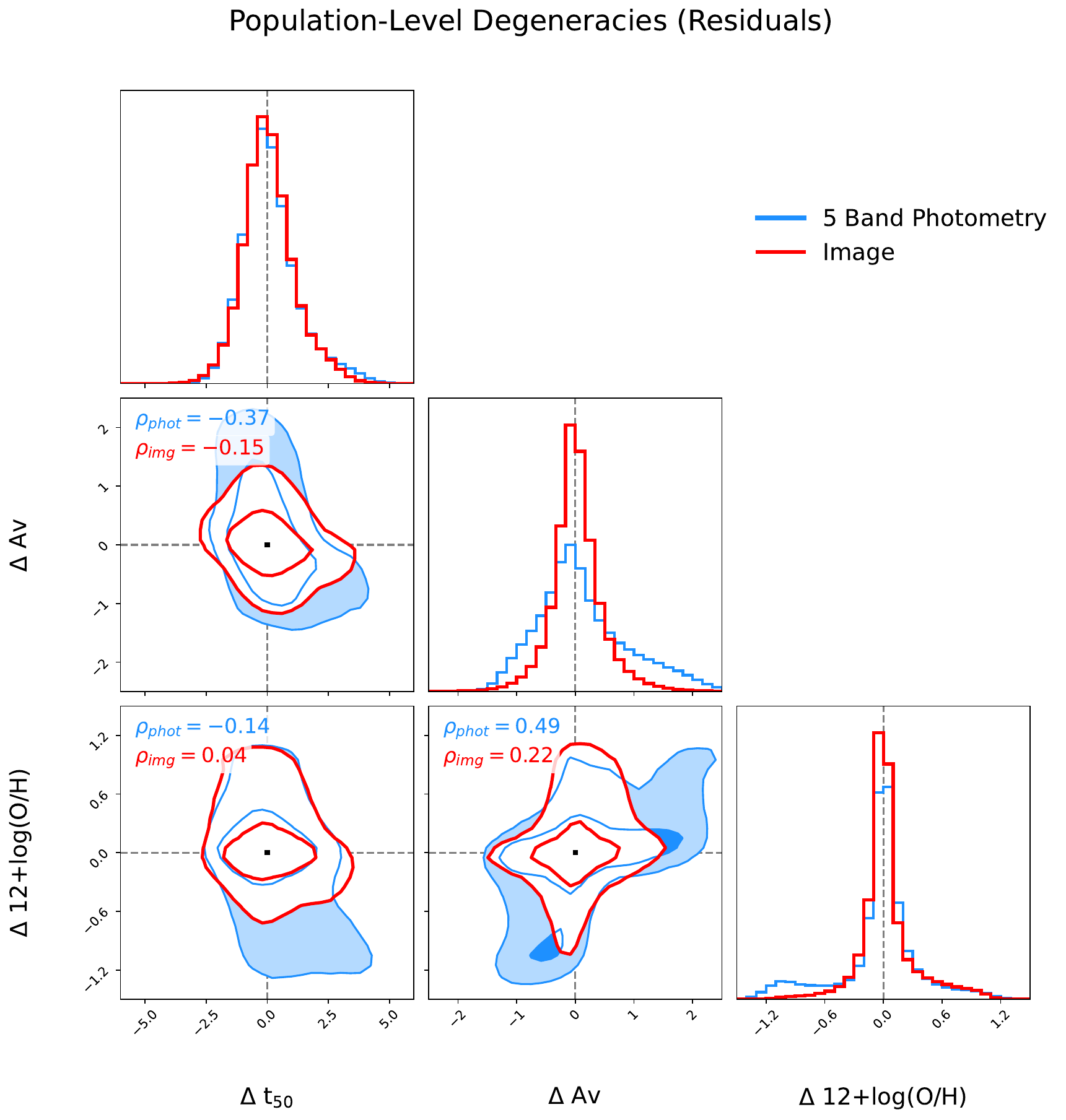}
    \caption{A population-level average inferred posteriors of t$_{50}$, Av, and 12+log(O/H) using the image model (red) and the 5-band photometry model (blue). The Spearman rank correlation coefficient ($\rho$) for each parameter pair, and for each model, is shown in the top left corner of each joint probability panel. The combined model reduces degeneracy compared to the 5-band photometry model in the traditionally hard-to-break age-dust-metallicity degeneracy. This suggests that key information to breaking the degeneracy is contained in the morphology of the galaxy.}
    \label{fig:pop_corner}
\end{figure}

The results for t$_{50}$, Av, and 12+log(O/H) are shown visually in \cref{fig:pop_corner} as a corner plot with the associated $\rho$ value for each pair. In \cref{fig:pop_corner} we show the centered residuals rather than the absolute parameter values to remove the intrinsic scatter of the population. By subtracting the median from each galaxy's posterior samples before pooling them, we center every galaxy's posterior on zero. This allows us to study the shapes of the produced posteriors without being biased by different galaxy populations. \cref{fig:pop_corner} shows that incorporating morphology substantially breaks the classical age--dust--metallicity degeneracy($\rho_{t_{50}-A_V}$ drops from $-0.37$ to $-0.15$, $\rho_{t_{50}-12+log(O/H)}$ drops from $-0.14$ to $0.04$, and $\rho_{A_V-12+log(O/H)}$ drops from $0.49$ to $0.22$). 

The results for each parameter pair across the imaging, combined, and 5-band photometry models respectively can be found in Appendix \ref{app:full_degen}. The imaging and combined models either reduce degeneracy between most parameter pairs or have similar degeneracy when compared to the 5-band photometry model.

\cref{fig:pop_corner} highlights that the image model significantly reduces degeneracy compared to the 5-band photometry model in the traditionally hard-to-break age-dust-metallicity degeneracy. While the combined model does a better job at breaking the degeneracies compared to the 5-band photometry model, it is only marginally better than the image model (for details see Appendix \ref{app:full_degen}). This suggests that key information for breaking degeneracy is contained in the morphology of the galaxy. The shape and structure of the galaxy help the model to separate the effects of age, dust, and metallicity on the observed photometry.

\subsection{Is Past Star Formation Encoded in Morphology?} \label{sec:sfh}
\begin{figure*}
    \centering
    \includegraphics[width=1\linewidth]{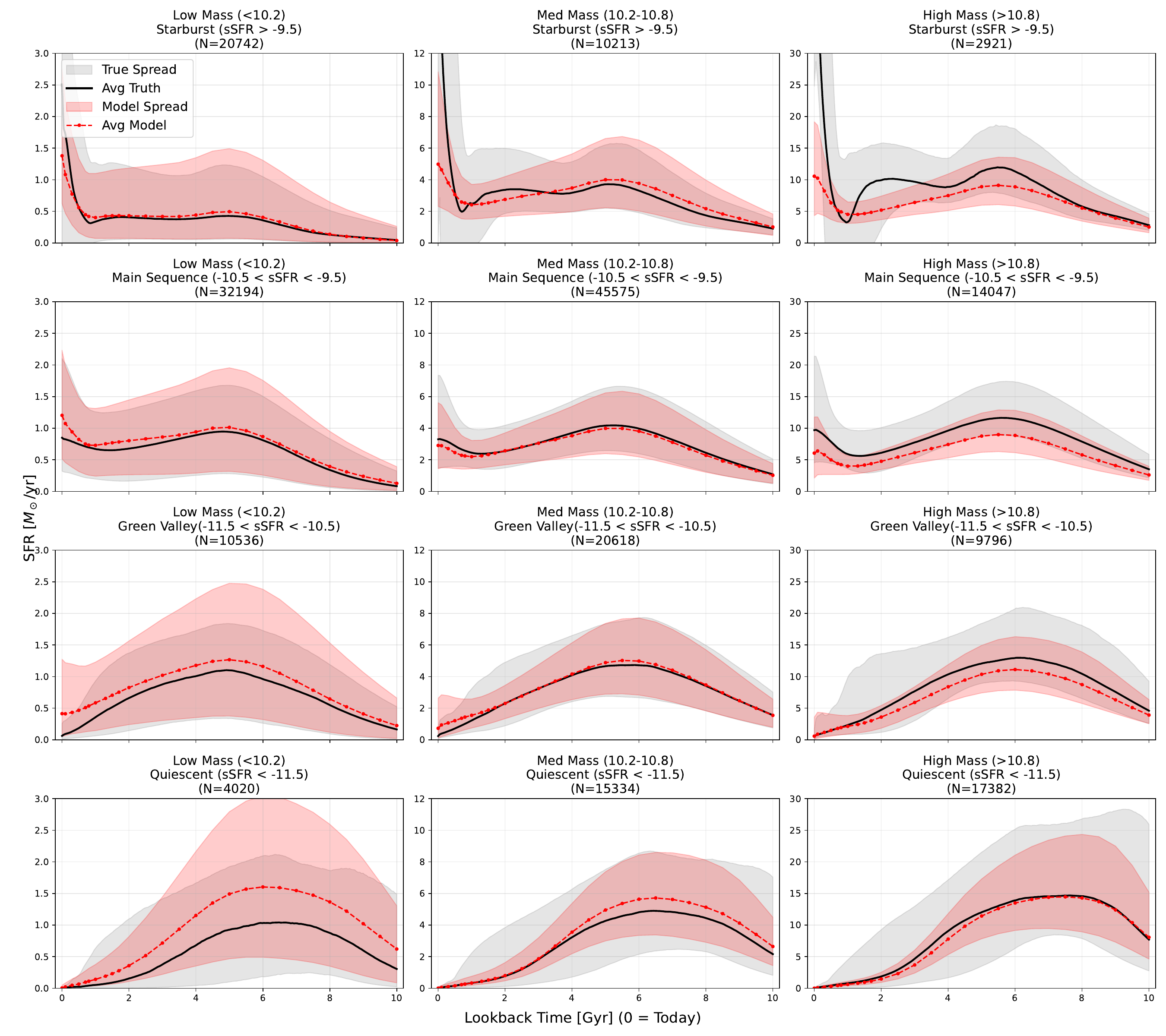}
    \caption{The accuracy of predicted SFHs in different parts of M$_*$-SFR space. Galaxies are divided into 3 mass bins and 4 sSFR bins. The average true SFH history for each bin is plotted in black, while the predicted SFHs are plotted in red. The 16th and 84th percentile sample dispersion for both the true and predicted SFHs is also plotted as the shaded regions around the SFHs. We find that on average for all galaxy populations, except starburst galaxies and green valley galaxies, we are able to recover the SFH all the way to 10 Gyr ago based solely on their present-day morphology and color with minimal error.}
    \label{fig:sfh_accuracy}
\end{figure*}

We explore how well image model is able to predict SFHs of galaxies. The accuracy of the predicted star formation histories, on a population level, can be seen in \cref{fig:sfh_accuracy}. Galaxies are divided into 3 mass bins and 4 sSFR bins in this plot to see how star formation reconstruction accuracy changes in different parts of the M$_*$-SFR plane. We use the following names and sSFR cutoffs for the four galaxy populations: starburst galaxies (sSFR $>$-9.5), main sequence galaxies (-10.5 $<$ sSFR $<$-9.5), green valley galaxies (-11.5 $<$ sSFR $<$-10.5), and quiescent galaxies (sSFR $<$-11.5). The shaded regions in this plot show the scatter in SFHs (both true values and predictions) within the given M$_*$ and sSFR bin. They do not show the confidence intervals of the median SFH in each bin. For some examples of predicted  SFHs with confidence intervals for individual galaxies, refer to Appendix \ref{app:indiv_sfh}.   

We find that on average for all galaxy populations, except starburst galaxies and green valley galaxies, we are able to recover the Dense Basis SFH conditioned on SED-fit summary statistics all the way to 10 Gyr ago based only on their present-day morphology and color with minimal error. A further exploration of why starburst and green valley galaxies have larger reconstruction errors is discussed in \cref{sec:sfh_discussion}. 

We aim to understand for how long the imprints of star formation remain on a galaxy's morphology. We investigate this idea using two approaches. We first see to what lookback time we can accurately predict SFR "continuously". Continuously here means that at every timestep up to a given lookback time we can accurately predict the SFR of the galaxy. For example, if we can continuously predict the SFR of a galaxy up to 1.0 Gyr ago, that means we could also accurately predict its SFR at 0.1 Gyr ago, 0.3 Gyr ago, 0.5 Gyr ago, 0.7 Gyr ago, and 1.0 Gyr ago. 

The second approach we consider is a "robust" prediction of SFR at a given lookback time. Robust here means we can accurately predict the SFR of the galaxy at some given lookback time (t) and the time step directly before it (t-1) and directly after it (t+1). However, we do not consider whether we could accurately predict SFR at intermediate time steps. For example, if we can robustly predict SFR at 1.0 Gyr ago, it does not necessarily mean we can also predict its SFR accurately at 0.1 Gyr ago, 0.3 Gyr ago, 0.5 Gyr ago, or 0.7 Gyr ago. However, it does mean that we can robustly predict SFR at 0.8 Gyr ago, 1 Gyr ago, and 1.3 Gyr ago. 

Since the SFH model uses an ILI framework, we are able to derive full posteriors of the predicted SFH. This means at each time step we have a full posterior of the galaxy's SFR at that lookback time. To define a prediction as "accurate," we ensure that the true SFR value at each time step lies within the 16th to 84th percentile confidence interval of the predicted posterior. 

The results of the continuous accuracy approach are shown in \cref{fig:cont_memory}. The results of the robust accuracy approach are shown in Appendix \ref{app:robust_memory}, as there is not as much variation in robust memory across galaxy populations as there is for continuous memory. 

 Starburst galaxies above the main sequence show very short continuous memories ($\approx$ 0-0.8 Gyr). Actively star-forming galaxies on the main sequence demonstrate notably long continuous memory timescales ($\approx$4-7 Gyr). Green valley galaxies just below the main sequence show short continuous memories ($\approx$2-4 Gyr). Quiescent populations lying well below the main sequence show longer continuous memory ($\approx$3-5 Gyr).

In contrast to their short continuous memory, starburst galaxies possess extended robust memory ($\approx$4-10 Gyr). Main sequence galaxies show long robust memories ($\approx$7-9 Gyr), complementary to their continuous memory. Green valley galaxies show much longer robust memories ($\approx$7-9 Gyr) compared to their continuous memories.  Quiescent galaxies display long robust memories ($\approx$8-10 Gyr). Overall, we find that all galaxy populations show longer robust memories than continuous memories, as their general long-term SFHs are traceable from their current-day morphology. 

A further discussion of the physical mechanisms driving the trends seen in continuous and robust memory can be found in \cref{sec:sfh_discussion}.

\begin{figure}
    \centering
    \includegraphics[width=1\linewidth]{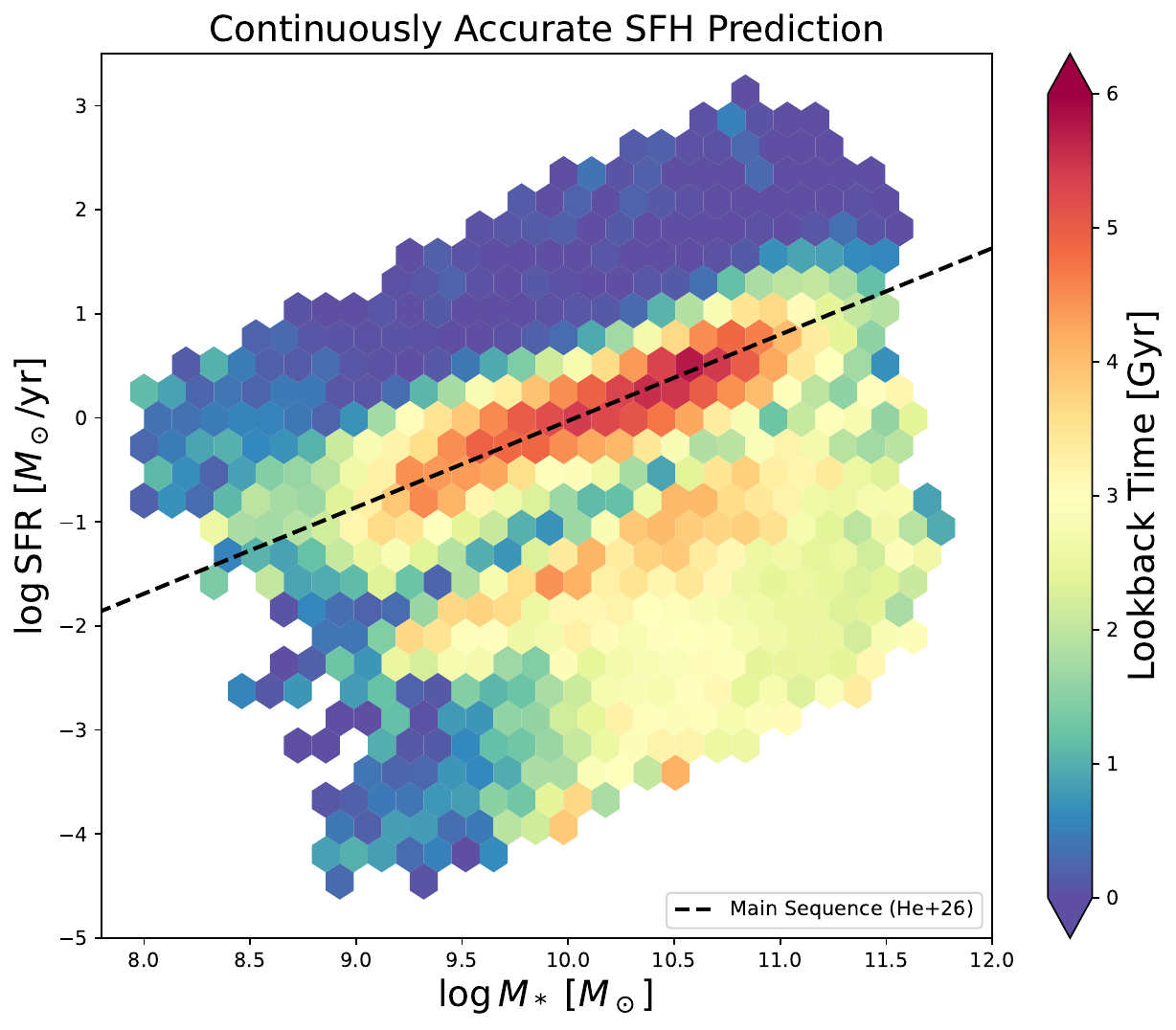}
    \caption{The continuous morphological memory of galaxies mapped across the M$_*$-SFR plane. The color scale denotes the continuous lookback time in Gyr over which the structural state of a galaxy remains an accurate record of SFH, defined by a 16–84$\%$ confidence interval. The star-forming main sequence from \cite{gswlc_sfms} is indicated by the dashed black line. Starburst galaxies above the main sequence show very short continuous memories ($\approx$ 0-0.8 Gyr). Actively star-forming galaxies on the main sequence demonstrate notably long continuous memory timescales ($\approx$4-7 Gyr). Green valley galaxies just below the main sequence show short continuous memories ($\approx$2-4 Gyr). Quiescent populations lying well below the main sequence show longer continuous memory ($\approx$3-5 Gyr).}
    \label{fig:cont_memory}
\end{figure}

\subsection{Interpreting Trends in Star Formation and Morphology} \label{sec:umap_results}
\begin{figure*}
    \centering
    \includegraphics[width=0.49\linewidth]{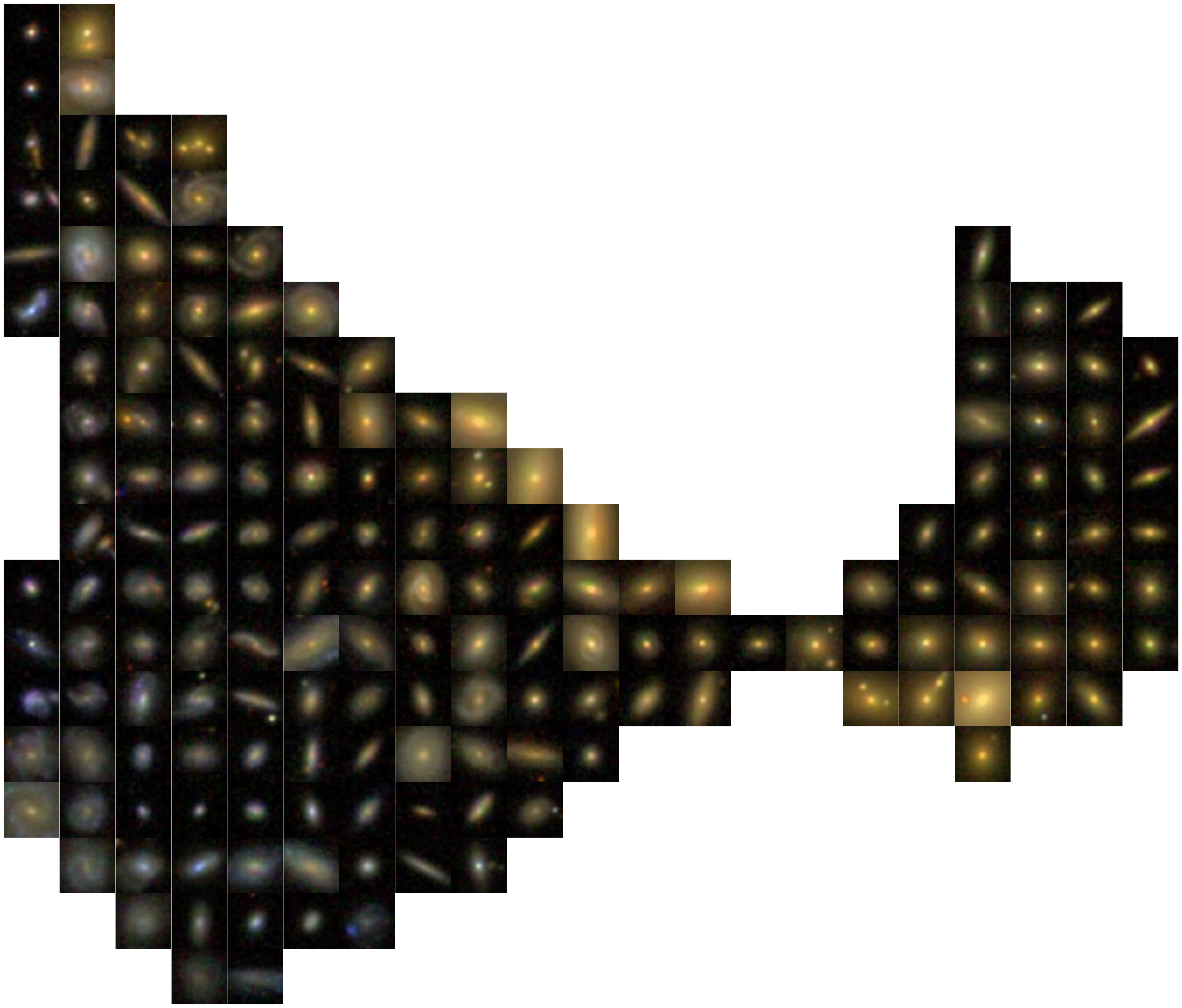}
    \includegraphics[width=0.49\linewidth]{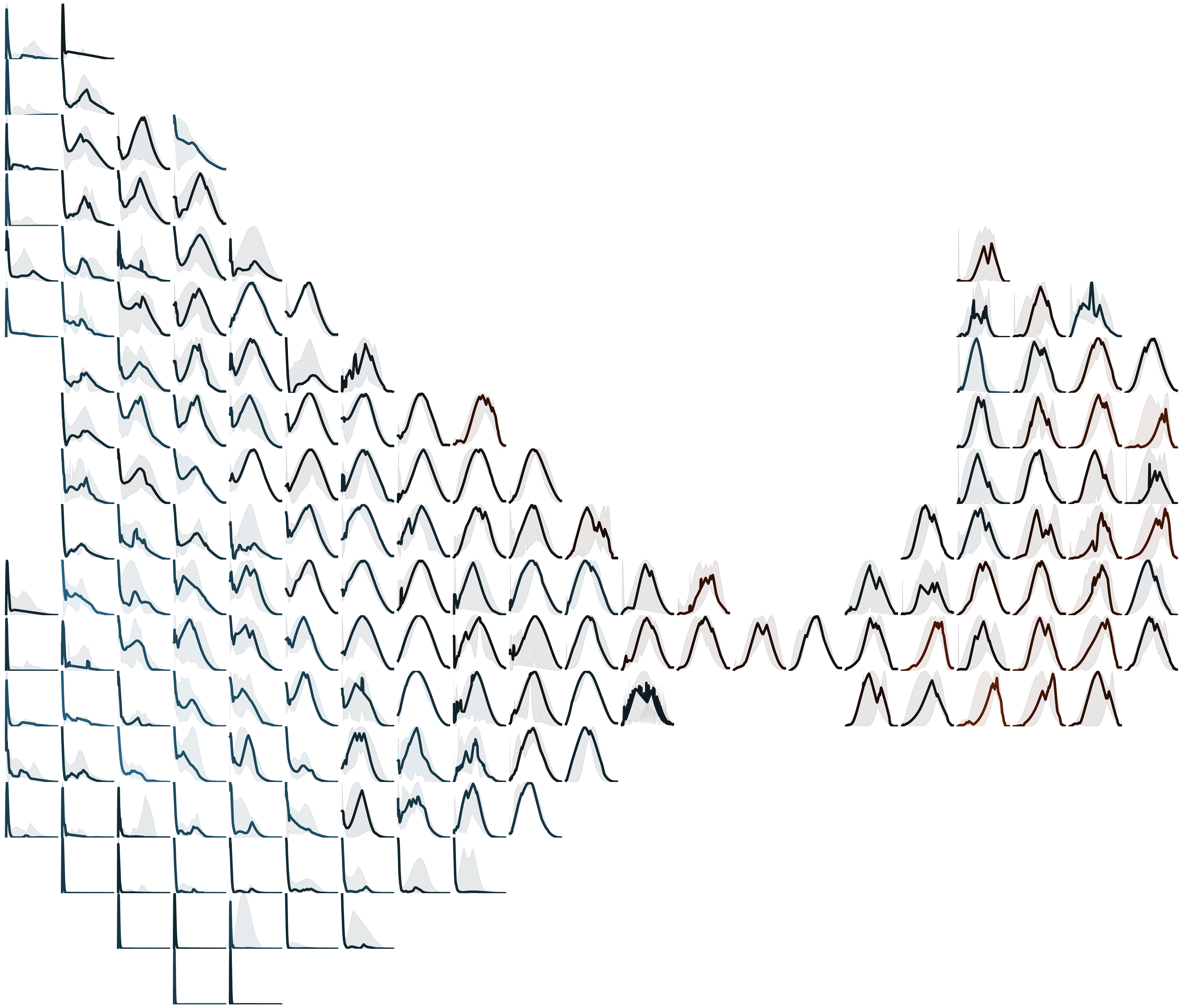}
    \includegraphics[width=1\linewidth, trim={0 0cm 0 0},clip]{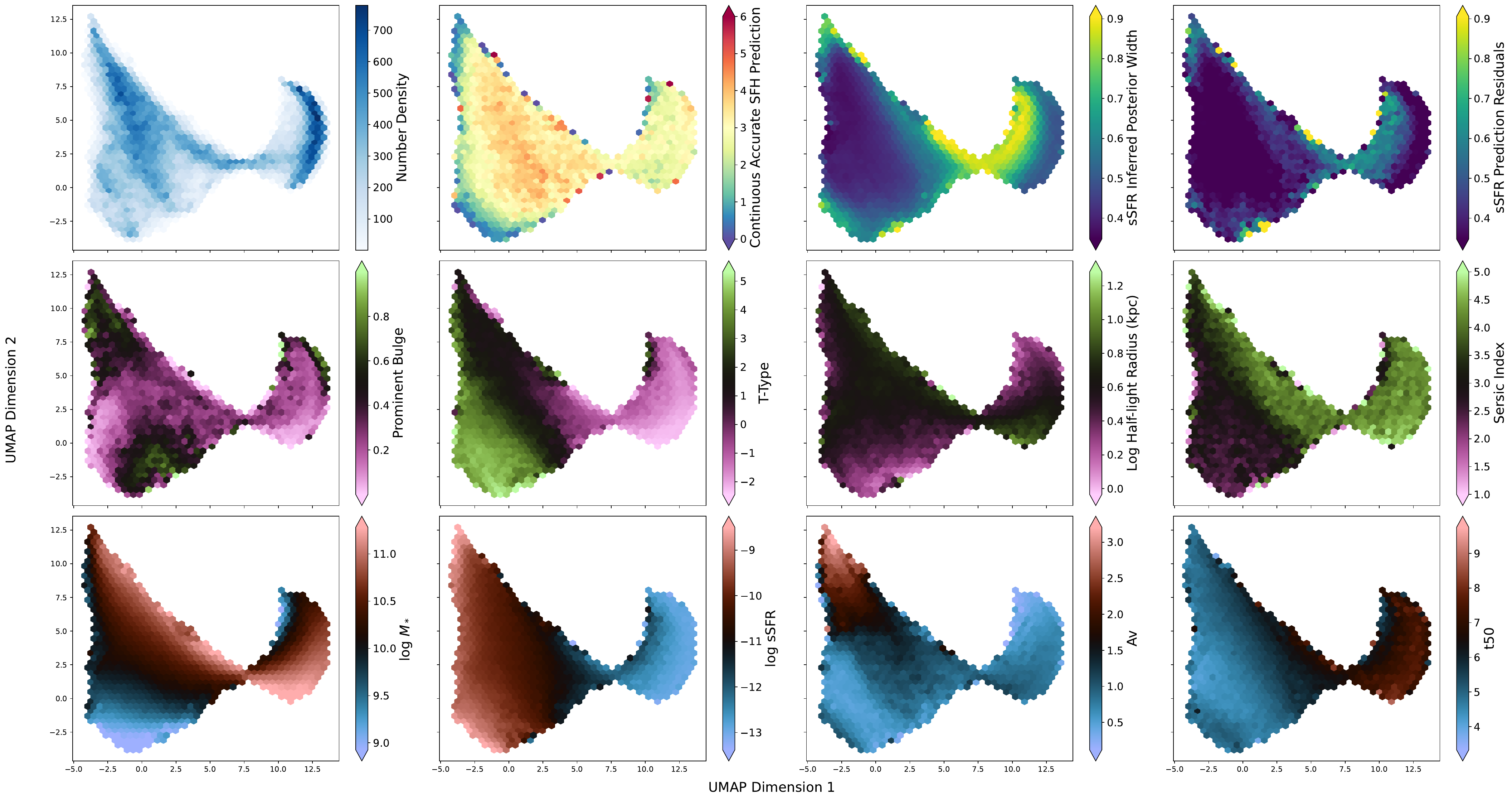}
    \caption{UMAP projection of the ViT image embedding space. The 2D projection reveals a highly structured, bimodal distribution that mirrors fundamental galaxy evolutionary sequences. The top left panel shows galaxy images from different parts of the UMAP space, while the top right one shows the median SFH of galaxies (in lookback time) in different parts of the UMAP space.  prediction metrics (sSFR residuals, sSFR inferred posterior width, continuous accurate SFH prediction), morphological properties (P$_{bulge}$, T-type, log of half-light radius, S\'ersic index), physical properties (log M$_*$, log sSFR, Av, t$_{50}$), and number density. The primary UMAP axis isolates quiescent, early-type galaxies (right cluster) from star-forming, disk-dominated galaxies (left cluster). Secondary gradients, such as the vertical dust (Av) sequence within the star-forming lobe, demonstrate that the ViT intrinsically links complex visual morphology to underlying physical states.}
    \label{fig:umap}
\end{figure*}

To see how effectively the ViT summarizes the morphology of the galaxies, we take a UMAP projection of the embedding space. The 2D projection of this space can be seen in \cref{fig:umap}. The plots in each panel are colored by: prediction metrics (sSFR residuals, sSFR inferred posterior width, continuous accurate SFH prediction), morphological properties (P$_{bulge}$, T-type, log of half-light radius, S\'ersic index), physical properties (log M$_*$, log sSFR, Av, t$_{50}$), and number density. 

\cref{fig:umap} shows a UMAP projection of the ViT image embeddings. A clear morphological sequence emerges across the primary UMAP dimension (horizontal axis). Galaxies with high bulge prominence ($P_{bulge} >0.8$) and early T-types cluster on the right, while disk-dominated galaxies ($P_{bulge} < 0.2$) cluster on the left. This morphological axis strongly correlates with physical properties. For instance, the plot colored by specific star formation rate (sSFR)  shows that the left side of the space (late-type galaxies) has systematically higher sSFR, while the right side (early-type galaxies) is quiescent. This can also be seen in the top right panel, which shows the SFHs of galaxies in different parts of the UMAP space. 

Furthermore, the embedding space captures secondary physical sequences within these broad morphological classes. The Av panel reveals a strong dust attenuation gradient entirely contained within the late-type, star-forming lobe, demonstrating the ViT's sensitivity to internal dust geometry rather than just global color.

Mapping the prediction metrics onto the embedding space reveals that the model’s precision is not uniform. The inferred posterior width for sSFR broadens at the extreme topological ends of the UMAP projection, indicating that the network is systematically less confident when processing the most extreme edge-case morphologies. This can be visually confirmed by looking at the top-left UMAP, which contains an image cutout of a random galaxy in different parts of the UMAP space. 

This visualization confirms the ViT has learned a meaningful representation that links visual structure to physical state.

\section{Discussion} \label{sec:discussion}

\subsection{How Much Information is Contained in Galaxy Morphology?}

\subsubsection{What is Known from Current Literature} \label{sec:morph_lit}

Different morphological features are intrinsically connected to a galaxy's SFH through distinct physical mechanisms.

Galactic disks contribute to stable, long-term star formation, which is well-described by Toomre-$Q$ parameter stability arguments and the gradual settling of gas disks over cosmic time \citep{disk,disk2,disk3,disk4}. 
In the absence of events like major mergers, these "settled disks" tend to exhibit relatively smooth star formation self-regulated by the disk \citep{disk_timescale,disk_timescale2}, which possibly contributes to the long-timescale memory of main-sequence galaxies we see in \cref{fig:cont_memory}. A majority of these tend to be disks and show a continuous memory qualitatively consistent with the times these disks are thought to have formed. 
 
Furthermore, the bulge-to-total mass ratio strongly correlates with a loss of angular momentum and subsequent quenching, often associated with mergers, feedback from the central supermassive black hole, or a combination of the two \citep{agn_bulge,agn_bulge2,agn_bulge3,agn_bulge4}. A massive bulge can also stabilize the remaining gas disk against further fragmentation, suppressing star formation without necessarily removing the gas. This would mean that the galaxy's star formation activity changes faster than their morphology, causing an overall reduction in their continuous memory until their morphological signatures "catch up".\citep{morph_quenching, morph_quenching3, Morph-class5, morph_quenching2, morph_quenching4}. 

Finally, galactic bars drive gas into the central regions to temporarily boost star formation, eventually becoming sites of quenching once the central gas reservoirs are exhausted \citep{bars,bars2,bars3,bars4}.

\subsubsection{Summary statistics of Morphology do not Capture Information About Past Star Formation}
Since classical summary statistics reduce complex, multi-scale structural data into single metrics, they inherently fail to capture the detailed SFH information embedded in the localized features discussed in \cref{sec:morph_lit}. While we do find that on the largest spatial and temporal scales there is a correlation between morphology and star formation activity (e.g. see the sSFR and T-type panels in Figure \ref{fig:umap}; \citealt{2015ApJ...811L..12W}), we find no strong correlation between the continuous memory and summary statistics of morphology such as: S\'ersic index ($\rho=-0.01$), T-Type ($\rho=0.02$), bulge-to-total ($\rho=-0.01$), and P$_{bulge}$ ($\rho=0.02$). This suggests that large-scale morphological features (which summary statistics describe) do not seem to encode sufficiently detailed information about past star formation activity. This suggests that information about long-term SFH might be present in the full 3-band images, which include smaller-scale morphological features (bars, clumps, tidal streams, etc.). 

Consequently, training implicit likelihood inference (ILI) models exclusively on these global morphological summary statistics does not allow for accurate SFH prediction. The loss of fine-scale structural data such as the presence of central bars, distinct clump distributions, or specific spiral arm topologies destroys the long-term fossil record of the galaxy's dynamical evolution. Full spatial imaging, rather than parameterized structural summaries, is strictly required to retain continuous morphological memory and accurately recover complex star formation histories. 

\subsubsection{Effects of Color and Image Size on Derived Properties} \label{sec:color}

As discussed in \cref{sec:pred_accuracy} and shown in \cref{tab:accuracy}, we find that the image model outperforms the 5-band photometry, similar to \cite{sed_morph}. The combined (image+5 band SED) model outperforms both the imaging and 5-band photometry model across all inferred parameters, and is actually comparable to the 11-band photometry. In this section, we discuss how much information is contained in galaxy morphology, and how this compares to extra information provided by additional photometric information.

We explore the extent to which the "extra information" comes from morphological structure vs color by running our analysis with monochromatic images (removing color information) and downsampled resolution (at half and quarter sizes). 

The bias and scatter of predicting each parameter in these tests are listed in \cref{tab:diff_images}, following the same format as \cref{tab:accuracy}. We see that the image model provides better precision overall, while the effects of color are more significant than those of image size. This is especially evident in heavily degenerate parameters like sSFR, age, dust, and metallicity, where color information heavily impacts the model's ability to be accurate. A similar result is shown in \cite{morph_v_color}, where color is shown to be more impactful than morphological structure when predicting the mass-to-light ratio of galaxies. 

The fact that we do not see a significant drop-off in scatter with image pixel size (1/2 vs 1/4 size in \cref{tab:diff_images}) suggests that as long as the model can make out the general \textit{spatial distribution of the light} rather than any specific morphological features, it can make adequate embeddings. Since these embeddings still accurately describe the general morphology of the galaxies, the ILI can make predictions comparable to when it has a larger image. 

\begin{table}[t]
    \centering
    \resizebox{\columnwidth}{!}{ 
    \begin{tabular}{llcc}
    \toprule
    \textbf{Model} & \textbf{Param} & \textbf{$\mu$} & \textbf{$\sigma_{\text{NMAD}}$} \\
    \midrule
    
    \multirow{6}{*}{\textit{gri} Image} 
    & $\log M_*$                & $\mathbf{0.000}$  & $\mathbf{0.154}$ \\
    & $\log \text{sSFR}$         & $\mathbf{0.028}$  & $\mathbf{0.438}$ \\
    & $z$                       & $\mathbf{-0.001}$  & $\mathbf{0.014}$ \\
    & $t_{50}$                  & $-0.030$           & $\mathbf{0.931}$ \\
    & $A_V$                     & $\mathbf{-0.004}$ & $\mathbf{0.303}$ \\
    & $12\!+\!\log(\text{O/H})$ & $\mathbf{0.000}$  & $\mathbf{0.124}$ \\
    
    \midrule
    \multirow{6}{*}{g-Band Image} 
    & $\log M_*$                & $-0.034$           & $0.325$ \\
    & $\log \text{sSFR}$         & $0.055$           & $0.725$ \\
    & $z$                       & $-0.002$           & $0.022$ \\
    & $t_{50}$                  & $-0.074$           & $1.051$ \\
    & $A_V$                     & $0.038$           & $0.540$ \\
    & $12\!+\!\log(\text{O/H})$ & $0.006$           & $0.156$ \\
    
    \midrule
    \multirow{6}{*}{Half-Size Image} 
    & $\log M_*$                & $0.007$           & $0.224$ \\
    & $\log \text{sSFR}$         & $0.043$           & $0.636$ \\
    & $z$                       & $\mathbf{0.001}$           & $0.020$ \\
    & $t_{50}$                  & $\mathbf{-0.012}$  & $0.966$ \\
    & $A_V$                     & $0.005$           & $0.437$ \\
    & $12\!+\!\log(\text{O/H})$ & $0.003$           & $0.153$ \\
    
    \midrule
    \multirow{6}{*}{Quarter-Size Image} 
    & $\log M_*$                & $0.009$           & $0.227$ \\
    & $\log \text{sSFR}$         & $0.043$           & $0.641$ \\
    & $z$                       & $0.001$           & $0.020$ \\
    & $t_{50}$                  & $-0.012$           & $0.958$ \\
    & $A_V$                     & $0.005$           & $0.432$ \\
    & $12\!+\!\log(\text{O/H})$ & $0.003$           & $0.153$ \\
    
    \bottomrule
    \end{tabular}
} 
\caption{Comparison of the bias and scatter (as described at the start of \cref{sec:pred_accuracy}) for the 4 different image ILI models. The \textit{gri} image model provides the lowest scatter in all predicted parameters. Color generally has a bigger effect on the scatter of the predicted parameters in comparison to image size.}
\label{tab:diff_images}
\end{table}

Additionally, the fact that adding imaging information to 11-band photometry (refer to Appendix \ref{app:accuracy}) did not significantly improve the performance of the model suggests that the 11-band photometry provides the maximal amount of information for predicting these labels at the level of accuracy our ground-truths allow. This makes sense considering these 11 bands form a superset of the 7 photometric bands used to derive the true values for training in this work. Since the target parameters were generated exclusively from a subset of this photometry, there is no additional morphological information for the model to extract that would improve its ability to reproduce these specific SED-fitted labels. This helps explain why the image model provides significant improvements to the 5-band photometry model, but not to the 11-band photometry model.

We also tested the role galaxy color plays in predicting the SFHs of galaxies. We find that both models recreate the general trends seen in \cref{fig:cont_memory}, but the 5 band model fails to recover some of the more subtle trends compared to the 11 band and image model. Compared to the 5-band photometry model, the image model recovers a more constrained relationship between high continuous memory and the SFMS. Compared to the 11-band photometry model and the imaging SFH model, the 5-band photometry SFH model does not find low continuous memory for green valley galaxies across a wide mass range, and does not find moderate continuous memory for quiescent galaxies (especially so for those around $10^{10.5}M_{\odot}$). Appendix \ref{app:color_sfh} contains more details on the continuous memories recovered from both modalities.
 
These results suggest that while a galaxy's color can act as a first order constrain to a galaxy's SFH, the galaxy's morphology provides complementary information that helps constrain the galaxy's SFH. Similar results of the effects of color vs morphology on SFH were found in \cite{katachi}. 

\subsection{The Connection Between Morphological Features and Physical Parameters}

\subsubsection{How does Star Formation Drive Structural Change?}
While morphology acts as an observable predictor for star formation, the physical processes of star formation and their associated feedback actively drive structural evolution in reverse. Sustained, steady star formation generally promotes inside-out growth and disk settling, steadily building angular momentum and forming extended structures \citep{disk,disk2,disk3,disk4}. Conversely, highly dynamic phases such as starbursts, post-starburst transitions, or stochastic events drive rapid structural change \citep{g_valley,g_valley2,g_valley3,starburst,starburst4,starburst2,starburst3}. 

The primary horizontal axis of the UMAP projection seen in \cref{fig:umap} reveals a distinct bimodal distribution that maps directly onto the physical transition associated with structural compaction and morphological quenching. The extended cluster on the left is heavily populated by star-forming, disk-dominated galaxies characterized by high specific star formation rates (sSFR), late T-types, and low bulge prominence ($P_{bulge}$). Sustained, steady star formation in these systems promotes continuous inside-out growth and dynamic disk settling over cosmic time \citep{disk,disk2,disk3,disk4}. 

Conversely, the concentrated cluster on the right isolates compact, quiescent, early-type galaxies with low sSFR and high $P_{bulge}$. This topological separation mirrors established literature on compaction: as galaxies undergo stochastic starbursts or violent disk instabilities, they experience rapid central mass build-up. This growth of a massive central bulge subsequently stabilizes the remaining gas disk against fragmentation, suppressing further large-scale star formation and driving the galaxy across the UMAP space into the quiescent cluster \citep{morph_quenching, morph_quenching3, Morph-class5, morph_quenching2, morph_quenching4}. 

Beyond this primary evolutionary axis, the UMAP embedding space captures secondary physical sequences within the star-forming lobe that highlight how different modes of star formation dictate structural change. The lower panels of \cref{fig:umap} reveal a pronounced vertical dust attenuation (Av) gradient entirely contained within the late-type cluster. This gradient separates steady, unobscured star-forming disks from highly stochastic, dust-enshrouded starburst galaxies. The ViT successfully distinguishes the regular spiral patterns of steady star formation from the disturbed, clumpy morphologies generated by violent, bursty star formation, placing them in distinct topological regions of the embedding space. 

Finally, the narrow structural "bridge" connecting the star-forming and quiescent clusters provides critical insight into transitional populations like green valley and post-starburst galaxies. The top-right panel of \cref{fig:umap}, which maps median star formation histories across the UMAP manifold, demonstrates that galaxies occupying this intermediate bridge exhibit rapidly declining recent star formation trajectories. This suggests that once a stochastic starburst exhausts its available gas reservoir or triggers energetic central feedback, the galaxy undergoes a rapid morphological transformation—shedding its star-forming clumps and asymmetric features—before settling permanently into the stable, early-type morphology seen on the far right of the projection.

\subsubsection{How Far Back in Time can Morphology "See"?} \label{sec:sfh_discussion}

Using the SFH model described in \cref{sec:sfh}, we aim to see how long imprints of star formation activity remain on a galaxy's morphology. In other words, we investigate how good of a "fossil record" present-day morphology is for a galaxy's SFH. We adopt the terms "continuous morphological memory" and "robust morphological memory" as outlined in \cref{sec:sfh}.

\cref{fig:sfh_accuracy} reveals our ability to, on average for all galaxy populations, except starburst galaxies and green valley galaxies, recover the SFH all the way to 10 Gyr ago based only on their present-day morphology and color with minimal error. The ability to trace the SFH of galaxies from their present-day morphology has been shown in previous works \citep{Morph_time6,katachi}. \cref{fig:sfh_accuracy} highlights that starburst galaxies and green valley galaxies show larger errors when predicting sSFR rates at shorter lookback times, likely due to being highly dynamic and active systems at the time of observation. For starburst galaxies this leads to their morphologies being more disturbed than other populations \citep{starburst,starburst4,starburst3,starburst2}, and for green valley galaxies this leads to their morphological change lagging behind their rapid quenching (i.e their morphological change happens on a much larger timescale than their quenching does) \citep{g_valley,g_valley3,g_valley2}. This makes inferring the recent past of both of these populations much more difficult. This is further reflected in \cref{fig:cont_memory}, where we see that both starburst and green valley galaxies have the shortest continuous memory, often failing the accuracy threshold within the first few time steps. However, in \cref{fig:robust_memory} we see that despite their recent past being chaotic, we are nevertheless able to recover their deep histories (i.e they have high robust memory). This highlights that their current morphology still retains features of the galaxies' long-term predictable evolutionary SFH track, which makes it possible to infer their progenitor star-forming state. 

\cref{fig:cont_memory} reveals that actively star-forming galaxies on the main sequence demonstrate notably long continuous memory timescales ($\sim$4-7 Gyr). This reflects their steady evolution along the main sequence, where their structural changes track predictably with their SFHs. On the other hand, quiescent populations lying below the main sequence exhibit shorter continuous memory ($\sim$3-5 Gyr). 

This follows what we expect given our current understanding of galaxy dynamics. Main sequence galaxies are dynamic systems due to their ongoing star formation. Continuous gas accretion, star formation feedback, disk instabilities, and minor mergers constantly reshape their short-term morphology \citep{sfms_scatter4,sfms_scatter2,sfms_scatter6,sfms_scatter,sfms_scatter3,sfms_scatter5,2024A&A...691A..56C,sfms_scatter7,gswlc_sfms}. Despite this, their steady evolution along the main sequence allows them to have relatively long continuous memories. However, this state of dynamic change due to ongoing star formation leads to their continuous memory being shorter than their robust memory.   

Once a galaxy quenches and drops below the main sequence, it enters a state of passive secular evolution. Without active starbursts or major disruptive events, its morphology (typically becoming an early-type/elliptical) is "locked in" and remains stable for much longer continuous stretches of time \citep{q_morph2, q_morph, q_morph3, q_morph8}. Additionally, the concept of morphological quenching, in which the growing bulge of a galaxy stabilizes the disk, hence suppressing star formation, is well documented in the literature \citep{morph_quenching, morph_quenching3,Morph-class5, morph_quenching2, morph_quenching4}. However, while their morphology is stable after the major quenching event, recent dry mergers and galaxy harassment can cause a discrepancy between morphology (affected by dry merger/galaxy harassment) and SFH (unaffected by dry merger/galaxy harassment \citep{sfh_trends}), leading to a lower continuous memory for some quiescent galaxies. In essence, these dry mergers and/or galaxy harassment could decouple the current morphological state of the galaxy from its SFH. High-mass quiescent galaxies are more likely to undergo dry mergers \citep{dmerger,dmerger3,dmerger4,dmerger5,sfh_trends,dmerger2}, meanwhile low-mass quiescent galaxies are more likely to be affected by galaxy harassment \citep{stripping3,stripping2,stripping4,stripping,stripping5,sfh_trends}. This is a possible explanation as to why quiescent galaxies show truncated memories when compared to main-sequence galaxies. 

Main-sequence galaxies possess highly extended robust memory ($\sim$8-10 Gyr), highlighting globally predictable long-term evolutionary tracks. Quiescent galaxies display a long robust memory ($\sim$8-10 Gyr); however, some of them have truncated robust memories. For these quiescent galaxies, this possibly reflects the drastic morphological change introduced by sudden and often violent quenching events.

While star-forming galaxies fluctuate continuously in their exact state (as seen in \cref{fig:cont_memory}), their overall evolutionary track is highly predictable \citep{Mass-SFR-Speagle, SFMS, Mass-SFR-Chang, Mass-SFR-Pan}. A galaxy currently evolving steadily on the main sequence has likely been growing predictably along that sequence for most of cosmic time. This is likely why the model can look back 8–10 Gyr and confidently infer its progenitor star-forming state.

For quiescent galaxies, we expect long robust memories since their morphology has been locked in once they stopped forming stars, and this is indeed the case for most of the population. However, some quiescent galaxies show shorter robust memories because quenching can be accompanied by a violent or rapid morphological transformation such as a major merger, compaction event, or strong AGN feedback \citep{q_morph9,q_morph5,q_morph7, q_morph8, q_morph10, q_morph6,q_morph4}. This violent transition significantly alters their morphology and acts as an information barrier to their deeper SFH. While their recent continuous memory is stable (as seen in \cref{fig:cont_memory}), trying to predict their deep past (before the quenching event) is much harder, limiting their robust memory. 
 
Taken together, continuous memory and robust memory illustrate the difference in timescales for which imprints of star formation are conserved in different galaxy groups. Starburst galaxies' morphologies and green valley galaxies' morphologies are both in a stage of transition from being in highly dynamical states. Despite this, however, their current morphology still retains features of the galaxies' long-term predictable evolutionary SFH track. Star-forming galaxies' morphologies are short-term stochastic, but retain long-lived morphological features from a stable evolution along the SFMS. Quenched galaxies' morphologies are mostly stable after their major quenching event, but morphological features present before the quenching event can be erased. Furthermore, quenched galaxies that undergo dry mergers or galaxy harassment could have their morphology altered, while their SFH remains mostly unchanged. Dry mergers and/or galaxy harassment in these systems could therefore decouple the current morphology of the galaxy from its recent SFH.

\subsubsection{Toy Model for Morphological Memory}
\label{sec:toy_model_formulation}

To understand whether the signatures we see in the continuous memory come from (i) decorrelation between morphology and SFHs due to violent events vs (ii) a difference between the rate of change of morphological signatures and star formation activity, we construct a phenomenological toy model that predicts the continuous memory for a population of mock SFHs under these two limiting physical regimes: an instantaneous structural response model ($t_1 \approx t_2$), where morphological transitions perfectly track fluctuations in the Star Formation Rate (SFR), and a morphological lag model ($t_1 \ll t_2$), where global structure relaxes on a timescale significantly longer than the transient gas-dynamical events driving the SFR.

We generate a synthetic sample of 15,000 star formation histories tracing standard Main Sequence (MS) secular evolutionary tracks, with physically motivated perturbations to generate representative subsets of Starburst (SB), Green Valley (GVG), and Quiescent (Q) systems. To accurately reflect observed demographics, the GVG population is explicitly subdivided into fast-quenching (55\%) and slow-quenching (45\%) trajectories \citep{gvg_quench,gvg_quench2,gvg_quench3}. These perturbed evolutionary tracks serve as the ground-truth star formation histories, $\mathrm{SFH}_{\mathrm{true}}$, against which our modeled predictions are evaluated.

In the delayed-response scenario, we emulate the outputs of our neural density estimator, $P(\theta_{\mathrm{sfh}} \mid \theta_{\mathrm{morph}})$, by recognizing that present-day morphology acts similar to a low-pass filter. We introduce a "morphological sensitivity" parameter, defined as the exponential decay of the timescale ratio, $\mathcal{L} = \exp(-t_2/t_1)$. This parameter dictates the exact fraction of a recent, violent star formation event that has successfully imprinted onto the galaxy's present-day observable structure. The inferred SFH is thus modeled as a weighted superposition of the true episodic history and the baseline secular MS track ($\mathrm{SFH}_{\mathrm{sec}}$):
$$ \mathrm{SFH}_{\mathrm{pred}} = (1 - \mathcal{L}) \cdot \mathrm{SFH}_{\mathrm{sec}} + \mathcal{L} \cdot \mathrm{SFH}_{\mathrm{true}} + \epsilon_{\mathrm{drift}} $$
where $\epsilon_{\mathrm{drift}}$ represents an arbitrary constant noise floor accounting for fundamental uncertainties in the underlying SFH-morphology mapping. In cases of rapid quenching or sudden bursts ($t_1 \ll t_2$), $\mathcal{L} \to 0$, forcing the network to anchor its recent predictions to the unperturbed secular track rather than the true episodic event.

Conversely, in the instantaneous regime, the network theoretically possesses immediate structural access to the perturbed state. We model the inferred SFH simply as:
$$ \mathrm{SFH}_{\mathrm{pred}} = \mathrm{SFH}_{\mathrm{true}} + \epsilon_{\mathrm{scramble}} $$
where $\epsilon_{\mathrm{scramble}}$ acts as a systematic scattering term proportional to the amplitude of the recent star-forming or quenching event, reflecting the informational disruption caused by the drastic morphological transition itself.

To turn these into a quantitative continuous memory metric, we compute the absolute residual error at each lookback timestep, $\Delta \mathrm{SFR}(t) = |\mathrm{SFH}_{\mathrm{pred}}(t) - \mathrm{SFH}_{\mathrm{true}}(t)|$. Similar to how we evaulate continuous memory, a predicted timestep is accurately recovered if the residual falls strictly below a defined fidelity threshold (here set to $0.3$ dex). The continuous memory is subsequently defined as the maximum contiguous lookback time, starting from $t_{\mathrm{obs}}$, over which the inferred SFH uninterruptedly satisfies this accuracy criterion before the first threshold failure occurs. This is the same as the continuous memory outlined in \cref{sec:sfh}. 

\begin{figure*}
    \centering
    \includegraphics[width=1\linewidth]{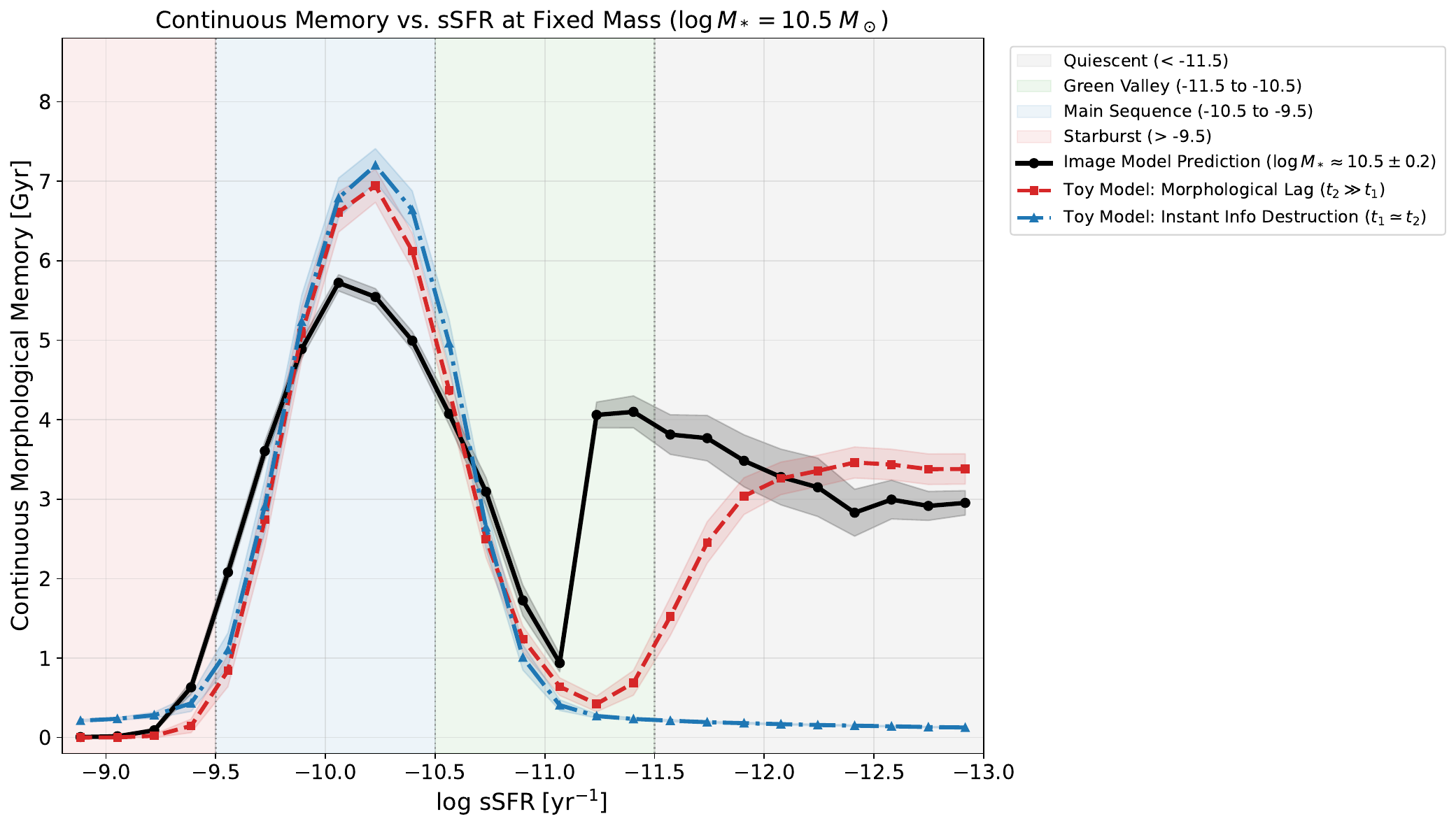}
    \caption{A comparison of continuous memory as a function of sSFR for predictions from the image model (black), the morphological lag toy model (red), and the instantaneous morphological change toy model (blue). The results are shown for a fixed mass bin of $10.3 < \mathrm{log}$ $M_* < 10.7$. The morphological lag toy model largely agrees with the image model predictions across all galaxy populations, while the instantaneous morphological change toy model agrees with the image model predictions for starburst and main sequence galaxies.}
    \label{fig:toy_model.pdf}
\end{figure*}

The results of the toy model compared to our image model predictions for a given mass bin of $10.3 < \mathrm{log}$ $M_* < 10.7$ can be seen in \cref{fig:toy_model.pdf}. We see that the delayed model closely aligns with the trends of continuous memory in our image model predictions, while the instantaneous model only aligns with trends in starburst galaxies and main sequence galaxies. 

\subsection{Improving Parameter Constraints with Morphology for Rubin/Roman/Euclid}
The idea that morphology contains unique information not found solely in photometry is not surprising. For example, a red and centrally concentrated galaxy has a very similar photometry to a more extended red galaxy with a disk. The photometry is unable to distinguish this disk feature, but we know a disk structure is extremely informative when trying to break something like the age-dust degeneracy. The presence of a disk, or perhaps the absence of a bulge, usually suggests that the galaxy is younger, as it has not yet undergone events like major mergers and cold gas accretion which build up its bulge \citep{1999ApJ...514...77N,2022MNRAS.510.1772N,2013MNRAS.433.2986W,2019MNRAS.488..609I}. Similarly, the presence of a disk, especially at higher masses, suggests that at least some star-forming activity remains, and we expect this to be associated with a dustier profile \citep{2005MNRAS.363..131P,2013MNRAS.432.2298B,2015ApJ...805..108P,2018A&A...615A...7L,2024ApJ...970..188N}. 

Given this, it seems that including morphology in our SED fitting codes for next-gen surveys is an ideal next step to take. It provides rich and unique information that can complement existing SED fitting codes. This has already been shown to be effective in recent works in which including galaxy images was shown to help reduce degeneracy in inferred parameters \citep{2025arXiv251205078Y,Yunus2026,morph_sed}.

Previous works have incorporated morphological information into SED fitting pipelines. However, these are mostly focused on improving photometric redshift predictions by incorporating summary statistic morphological information into the pipeline \citep{sed_morph2, sed_morph,2026MNRAS.548ag533R}. The improvements seen from including full morphological information (i.e the full image as is done in this work) rather than just morphological summary statistics are significant, as shown in various recent works \citep{2025arXiv251205078Y,Yunus2026,2026arXiv260210207K}. This once again highlights that there is significant loss of information when we reduce the morphology of a galaxy into a single number. In order to improve future SED fitting, we must then look for solutions that bypass the need to summarize the morphology into one number, and instead turn to a more holistic approach that incorporates the full galaxy structure/image.

The results shown in \cref{sec:pred_accuracy} suggest that for surveys where we only have access to a few photometric bands, especially if these bands are only in one part of the EM spectrum, including full morphological information can significantly help constrain the prediction of physical properties.  This makes the \katachiprime method ideal for making the most of the data from next-gen large-scale galaxy surveys from Euclid, Roman and Rubin. These surveys will have millions of sources observed, with the vast majority only having photometric broadband coverage.

The Euclid Space Telescope utilizes a dual-instrument suite comprising the Visible Instrument (VIS) and the Near-Infrared Spectrometer and Photometer (NISP). VIS operates with a single broad visible band spanning 550 to 900 nm and achieves a spatial resolution of 0.1 arcseconds per pixel. Complementing this optical capability, NISP provides photometric coverage across three near-infrared filters: Y, J, and H. They collectively span a wavelength range from 920 to 2000 nm—with a coarser pixel scale of 0.3 arcseconds per pixel \citep{euclid_VIS,euclid}. For image-based inference frameworks like \katachiprime, Euclid's leap in spatial resolution—an order of magnitude sharper than standard ground-based SDSS imaging—will allow neural architectures to decode much finer morphological imprints of past star formation, such as subtle stellar clumps or tidal features. Furthermore, the addition of deep near-infrared photometry will provide crucial constraints on older stellar populations, significantly improving the accuracy of spectral energy distribution (SED) and star formation history (SFH) predictions at higher redshifts compared to our current datasets.

Operating alongside Euclid in the space-based domain, the Nancy Grace Roman Space Telescope features the Wide Field Instrument (WFI), which offers a field of view roughly two orders of magnitude larger than that of the Hubble Space Telescope’s infrared channels. The WFI is equipped with eight overlapping filters covering the visible to near-infrared spectrum (0.48 µm to 2.3 µm). Roman achieves high spatial resolution with a focal plane pixel scale of 0.11 arcseconds per pixel, yielding an expected Point Spread Function Full Width at Half Maximum that ranges from approximately 0.058 arcseconds at the shortest wavelengths to 0.15 arcseconds at the longest \citep{roman}. By densely sampling the optical-to-NIR spectrum at such high spatial fidelity, Roman data will effectively act as spatially resolved SEDs. Feeding these multi-band, high-resolution images into our vision architecture is expected to dramatically decrease the uncertainty in predicted physical properties, enabling highly accurate SFH and stellar mass mapping across vast survey volumes.

In contrast to these space-based observatories, the Vera C. Rubin Observatory in Chile will execute the Legacy Survey of Space and Time (LSST) from the ground using a 3.2-gigapixel camera. The observatory employs six broad optical to near-infrared filters (u, g, r, i, z, and y), which span an overall wavelength range of 320 to 1050 nm. Although the camera boasts a native pixel scale of 0.2 arcseconds per pixel, the effective spatial resolution of the photometric data is fundamentally constrained by Earth's atmosphere, with the Cerro Pachón site exhibiting an expected median atmospheric seeing of approximately 0.67 arcseconds \citep{rubin}. Despite this atmospheric limit, LSST's images will be sharper and considerably deeper than legacy SDSS imaging. This unprecedented depth and multi-band optical coverage will yield a massive statistical volume, enabling wide-area science that was previously impossible. For example, tracking morphological transformations and quenching mechanisms across diverse cosmic environments for billions of sources.

To best implement the \katachiprime method on data from these next-generation surveys, it will be critical to use complementary data from legacy telescopes to build a robust "golden sample." In this work, we use a synergy of GALEX, SDSS, and WISE data to derive the most reliable physical parameters possible for use as training labels. Moving forward, existing fields with extensive multi-wavelength coverage or full spectroscopy (e.g., COSMOS or GOODS) can similarly be leveraged to establish highly robust training sets for Euclid, Roman, and Rubin. The trained method can then be deployed across the full footprint of these wide-area next-gen surveys. This strategy allows us to bypass the need for complete spectral coverage for every object, rapidly generating robust predictions for galaxy properties based solely on the new survey photometry.

\subsection{Caveats and Limitations} \label{sec:limitations}
A primary limitation of utilizing vision transformers for astrophysical inference is the challenge of disentangling whether the network is learning fundamental, generalizable physical mappings or simply keying into specific priors and biases intrinsic to the training dataset. Since neural networks are highly efficient pattern optimizers, they are prone to leveraging secondary observational artifacts or sample distributions as predictive shortcuts rather than extracting true physical information from a galaxy's structure.

 For example, the apparent angular size and surface brightness of a galaxy predictably diminish with distance. The vision transformer may heavily weight these superficial scaling factors to indirectly infer redshift or stellar mass. Rather than identifying scale-invariant structural details like spiral arm pitch angles or central bars, the model might simply learn that "smaller and dimmer equals further away." If the network overly relies on these apparent scaling relations, its predictive power becomes brittle and tightly coupled to the specific magnitude limits and selection functions of the SDSS and GSWLC datasets. This is especially important when considering datasets that span a much larger range of redshift than what is used in this study. 

 Additionally, the model could overfit to the underlying population distributions present in the local universe sample ($z < 0.16$). If the network learns that red, quiescent galaxies statistically dominate at higher masses within this specific training set, it may default to predicting low specific star formation rates based entirely on broad global color gradients. In this scenario, the network functionally ignores the fine-scale morphological structures that \katachiprime is designed to exploit, operating instead as a highly complex color-magnitude interpolator (which we discuss in \cref{sec:color}). Furthermore, future work is needed to understand how the performance of \katachiprime will scale at higher redshift, where galaxy images are less resolved, and hence contain less morphological information.
 
 Quantifying how much the network relies on these dataset-specific heuristics versus genuine morphological indicators remains a critical hurdle before deploying such methods to deeper, next-generation surveys where these underlying observational priors will drastically shift. This also highlights the importance of creating a representative training dataset that fully encapsulates the diversity of the full dataset we are interested in.   

Similarly, because our model relies on traditional SED fitting to generate the training labels, they are inherently limited by the lack of absolute ground truth. The ILI model may inadvertently inherit and reproduce the classical degeneracies, simplified star formation history parameterizations, and systematic uncertainties built into the underlying SED models. In its current state, what \katachiprime infers is a "SED-fit estimate" and not a physical "true value". To address this fundamental limitation in future work, \katachiprime must be trained and validated on synthetic observations generated from cosmological simulations that are properly forward modeled to match real observations. By utilizing simulations, where the intrinsic physical parameters and exact star formation histories are known precisely, we can circumvent the biases inherited from classical SED fitting routines and robustly test the network's predictive capabilities against a true physical baseline.

\section{Conclusion and Future Prospects} \label{sec:conclusion}
We introduced \katachiprime, a ViT and ILI-based method that can infer the physical properties and SFHs of galaxies from \textit{gri} broadband images with higher precision than 5-band optical photometry. \katachiprime leverages the unique information present in galaxy morphology in order to constrain physical parameter inference. 

The key advantages of the \katachiprime method over traditional SED fitting are:

\begin{enumerate}
    \item Can outperform photometry in surveys where there is a limited number of photometric bands
    \item Can use the unique data in the image/morphology of the galaxy 
    \item Can infer physical properties of galaxies much faster than traditional SED fitting (taking into account the training time of ILI)
    \item Breaks predicted parameter degeneracy better than traditional photometric SED fitting (given a comparable number of bands)
\end{enumerate}

The image model significantly outperforms 5-band photometry by identifying structural features, such as prominent bulges with associated disk structures and inclination angles (e.g., edge-on orientations). This spatial and structural information acts as a powerful non-photometric prior that pure photometry cannot access.

Furthermore, we investigated how far back in time morphology acts as a "fossil record" of a galaxy's star formation history. By modeling SFHs, we uncovered a stark dichotomy between a galaxy's "continuous" and "robust" morphological memory. Actively star-forming main sequence galaxies exhibit stable continuous memory ($\sim$4-7 Gyr) because their structural changes track predictably with their star formation. 

In contrast, highly dynamic populations like starburst and green valley galaxies have shorter continuous memories. For starburst galaxies this is due to disturbed morphologies from star formation activity. For green valley galaxies this is due to their morphological change happening on a longer timescale than their rapid quenching (this is discussed in detail in \cref{sec:toy_model_formulation}). However, despite this short-term stochasticity, starbursts and green valley galaxies retain an extended robust memory ($\sim$4-10 Gyr), preserving features of their long-term evolutionary tracks. 

Conversely, quiescent galaxies show moderate continuous memories ($\sim$3-5 Gyr) despite having mostly stable morphologies after quenching. This is likely due to dry mergers and/or galaxy harassment in these systems decoupling the current morphology of the galaxy from its recent SFH. Quiescent galaxies display a extended robust memory ($\sim$8-10 Gyr), with their morphology being locked in following violent quenching events. However, their quenching event can act as an information barrier for some quiescent galaxies, which erases signatures of their deep star formation past and makes them have shorter robust memories.

The \katachiprime method will be critical for ongoing and future large-scale surveys coming from telescopes such as Euclid, Roman, and Rubin. For these surveys, we expect to get mostly photometric data in only one part of the EM spectrum for an unprecedented number of galaxies. For this type of data, we have shown that \katachiprime is both more accurate and more computationally efficient than traditional methods. Leveraging the morphology of the observed targets, combined with using ILI to speed up the inference process, will allow us to make the most out of this data to explore the evolution of galaxies. 

In upcoming studies, we aim to expand this method to newly released data from Euclid. Euclid provides 1 visible very broadband photometry along with 3 NIR broadband photometry. With traditional SED fitting methods, this leaves us with little data to make constrained predictions of the physical properties of the observed galaxies. The \katachiprime method has shown to be better at constraining predictions of galaxy physical parameters when the available number of photometric bands is 5 or less and when they are restricted to a narrow range of the EM spectrum. The \katachiprime method, in conjunction with ancillary data from ground based surveys, will allow us to derive high-precision physical parameters of these galaxies with limited data and in a fraction of the time compared to traditional SED fitting methods. 

\begin{acknowledgments}
Support for J.P.A. was provided by the Japan Society for the Promotion of Science (JSPS) Core-to-Core Program (grant number: JPJSCCA20210003), the GPPU program at Tohoku University. Support for J.P.A. and N.S.H. was provided by the Japanese Ministry of Education, Culture, Sports, Science and Technology (MEXT) scholarship. J.P.A acknowledges the support of the Center for Computational Astrophysics at the Flatiron Institute and Columbia University for hosting him as a guest researcher during the initial stages of this work. Support for K.G.I. was provided by NASA through the NASA Hubble Fellowship grant HST-HF2-51508 awarded by the Space Telescope Science Institute, which is operated by the Association of Universities for Research in Astronomy, Inc., for NASA, under contract NAS5-26555.
\end{acknowledgments}

\software{
dense basis \citep{dense_basis,dense_basis_2}
ltu-ili \citep{ltu-ili}
matplotlib \citep{matplotlib}
Numpy \citep{numpy}
PyTorch \citep{Pytorch}
Astropy \citep{astropy1,astropy2,astropy3}
SciPy \citep{scipy}
hickle \citep{hickle}
FSPS \citep{FSPS,FSPS2}
sbi \citep{sbi}
Gemini \citep{gemini}
transformers \citep{transformers}
UMAP \citep{UMAP_lib}
corner \citep{corner}
pathfinder \citep{pathfinder}
          }

\bibliography{sample631}{}

\bibliographystyle{aasjournalv7}

\appendix

\section{Comparison to \katachi Method} \label{app:katachi}
As outlined in \cite{katachi}, the \katachi method was trained to predict 3 key properties of galaxies: M$_*$, sSFR, and t$_{50}$. With these 3 parameters, the star formation history of the galaxies could be reconstructed using dense basis \citep{dense_basis}. 

We compare how well the \katachi method is able to predict these 3 parameters for the GSWLC sample outlined in this paper. The results can be seen in \cref{tab:katachi vs katachi prime}, where we compare \katachi to the image model presented in this paper. \katachi does significantly worse in this sample when compared to \katachiprime. We suspect this is because the wide redshift range(0.02 $\textless$ z $\textless$ 0.16, median of z$=$0.08) present in this sample makes the prediction of stellar mass more difficult for \katachi, which was trained on a narrower redshift range (0.02 $\textless$ z $\textless$ 0.10, median of z$=$0.04). Without being given this redshift information, it struggles to understand the luminosity and distance relationship for these galaxies, leading to wrong predictions. We believe \katachiprime is able to bypass this issue, as the multidimensional posterior it builds for galaxy properties includes a redshift prediction, which helps it to understand the luminosity and distance relationship.

\begin{table}[h!]
    \centering
    \begin{tabular}{|c|c|c|}\hline
        Parameters  & \katachiprime  & \katachi \\\hline
         M$_*$ &  0.14 dex& 0.63 dex\\\hline
         sSFR &  0.32 dex& 3.30 dex\\ \hline
 t$_{50}$& 0.81 Gyr&4.63 Gyr\\\hline
    \end{tabular}
    \caption{Comparison of the prediction error for \katachi and \katachiprime on the GSWLC sample described in \cref{sec:data}. \katachiprime performs much better than \katachi, likely due to the fact that \katachi lacks redshift information.}
    \label{tab:katachi vs katachi prime}
\end{table}

\section{Random Test Set Galaxies Inferred Posteriors}
\label{app:indv_posteriors}
The posteriors inferred by the image model for 6 predicted physical parameters of a randomly selected sample of galaxies from the test set.
\begin{figure*}
    \centering
    \includegraphics[width=1\linewidth]{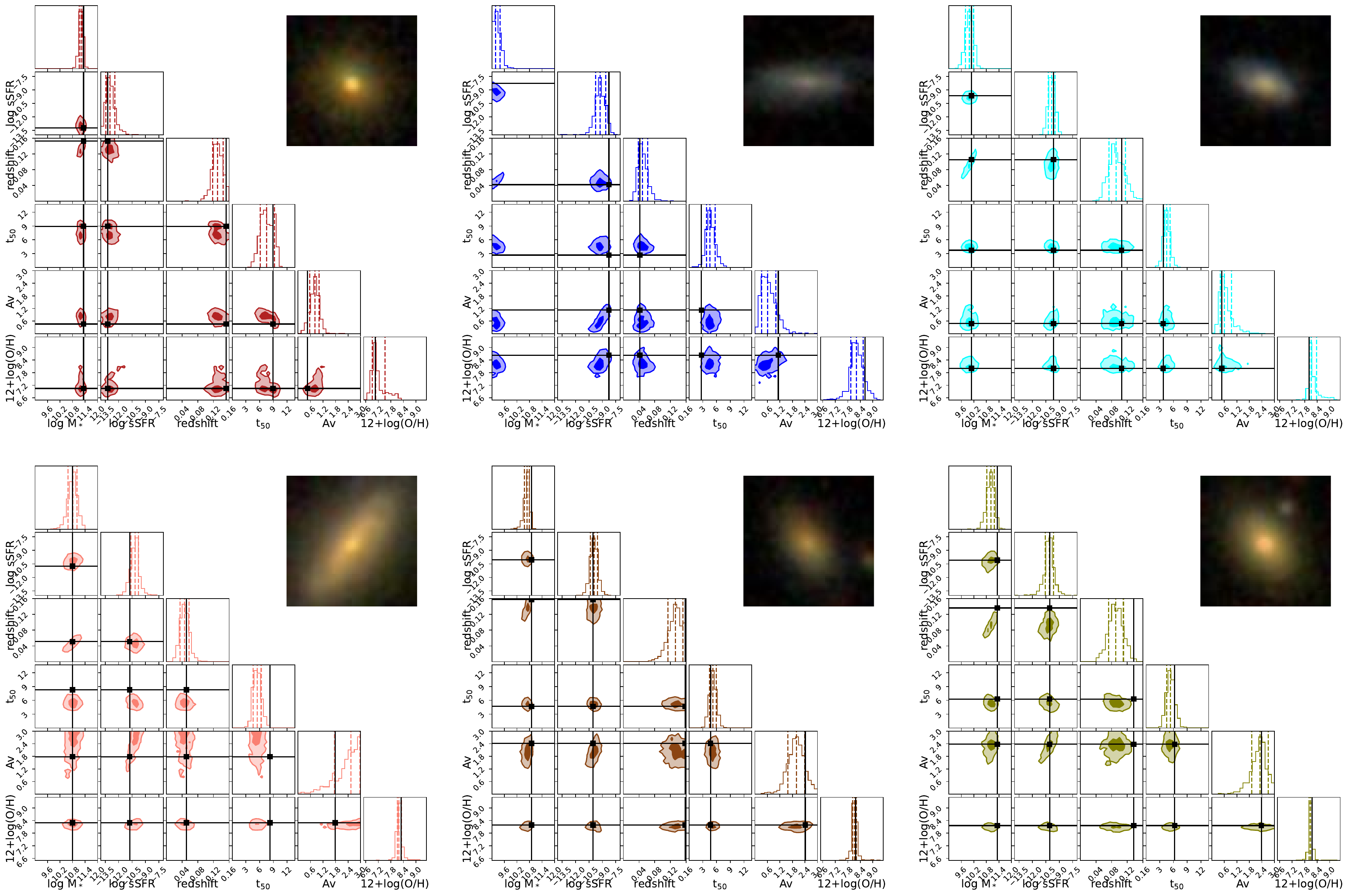}
    \caption{The posteriors inferred by the image model for 6 predicted physical parameters of a randomly selected sample of galaxies from the test set. Each galaxy was chosen randomly from one of 6 subdivisions in parameter space to show a variety of galaxy populations.}
    \label{fig:indv_corner_plots}
\end{figure*}

\section{Precision of 11-band Photometry} \label{app:accuracy}

\begin{figure*}
    \centering
    \includegraphics[width=0.85\linewidth]{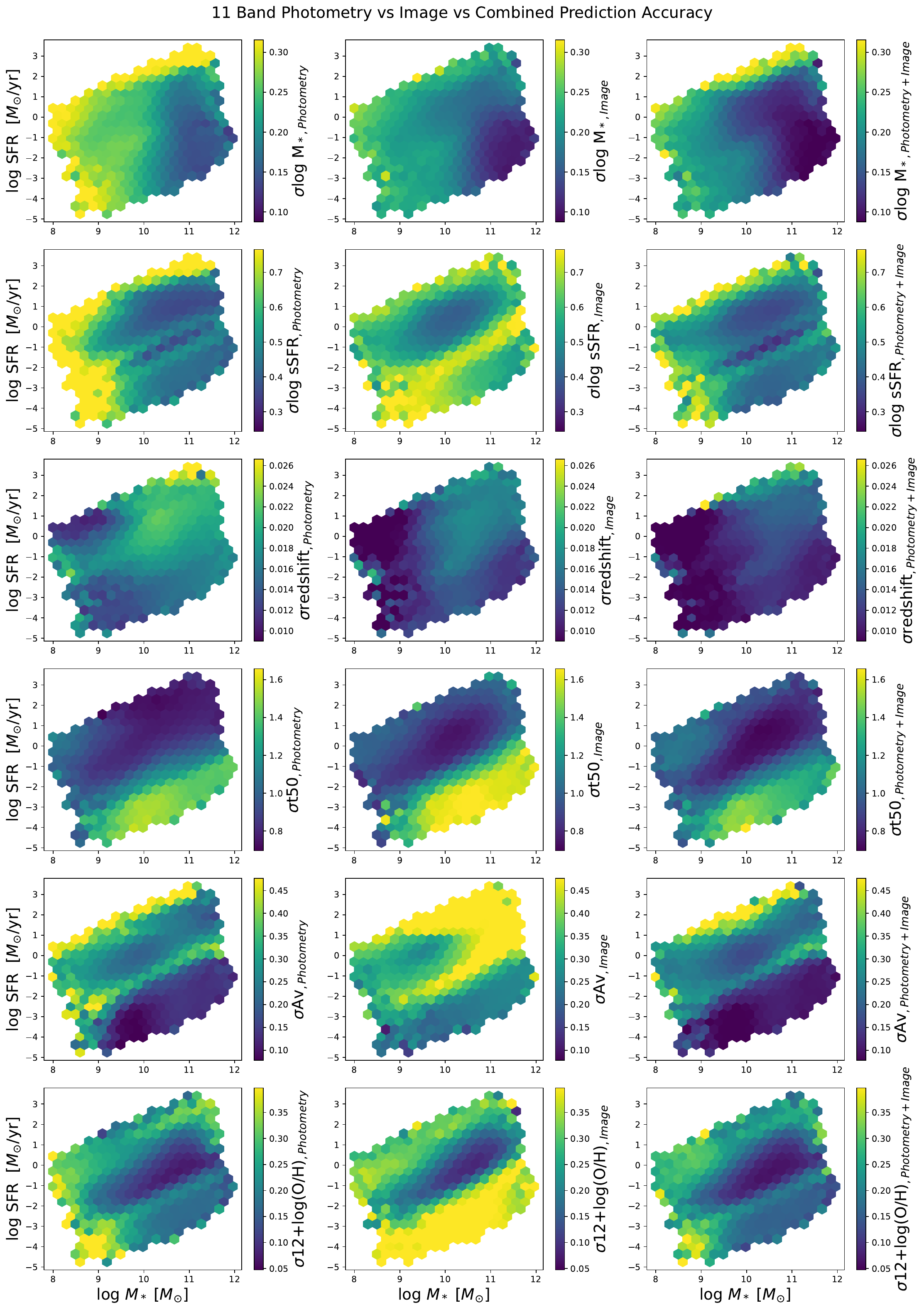}
    \caption{The prediction precision of the 11-band photometric model vs the image model vs the combined model in relation to the galaxies' location on the M$_*$-SFR plane. The first column shows the precision for the 5-band photometric model, the second column shows the precision for the image model, and the third column shows the precision of the combined model. Each row corresponds to one of the 6 inferred parameters (M$_*$, sSFR, z, t$_{50}$, Av, 12+log(O/H)). Adding morphological information to 11-band photometry is much less significant to precision than for 5-band photometry.}
    \label{fig:11_band_accuracy}
\end{figure*}

A comparison of how precise the image model is compared to the 11-band photometry model and the combined model (11-band photometry and imaging data) can be seen in \cref{fig:11_band_accuracy}. In \cref{fig:11_band_accuracy}, the galaxies are plotted on the M$_*$-SFR plane and are colored by the precision of predicting their physical parameters. Each of the 6 rows in the plot represents a different physical parameter, with the first column showing results for the 11-band photometry model, the second column showing results for the image model, and the third column showing the results for the combined model. Adding morphological information to 11-band photometry is much less significant to precision than for 5-band photometry. 

\section{11-band Photometry and Imagining Posteriors} \label{app:corner}

\begin{figure}
    \centering
    \includegraphics[width=1\linewidth,trim = 5cm 0cm 5cm 0cm,clip]{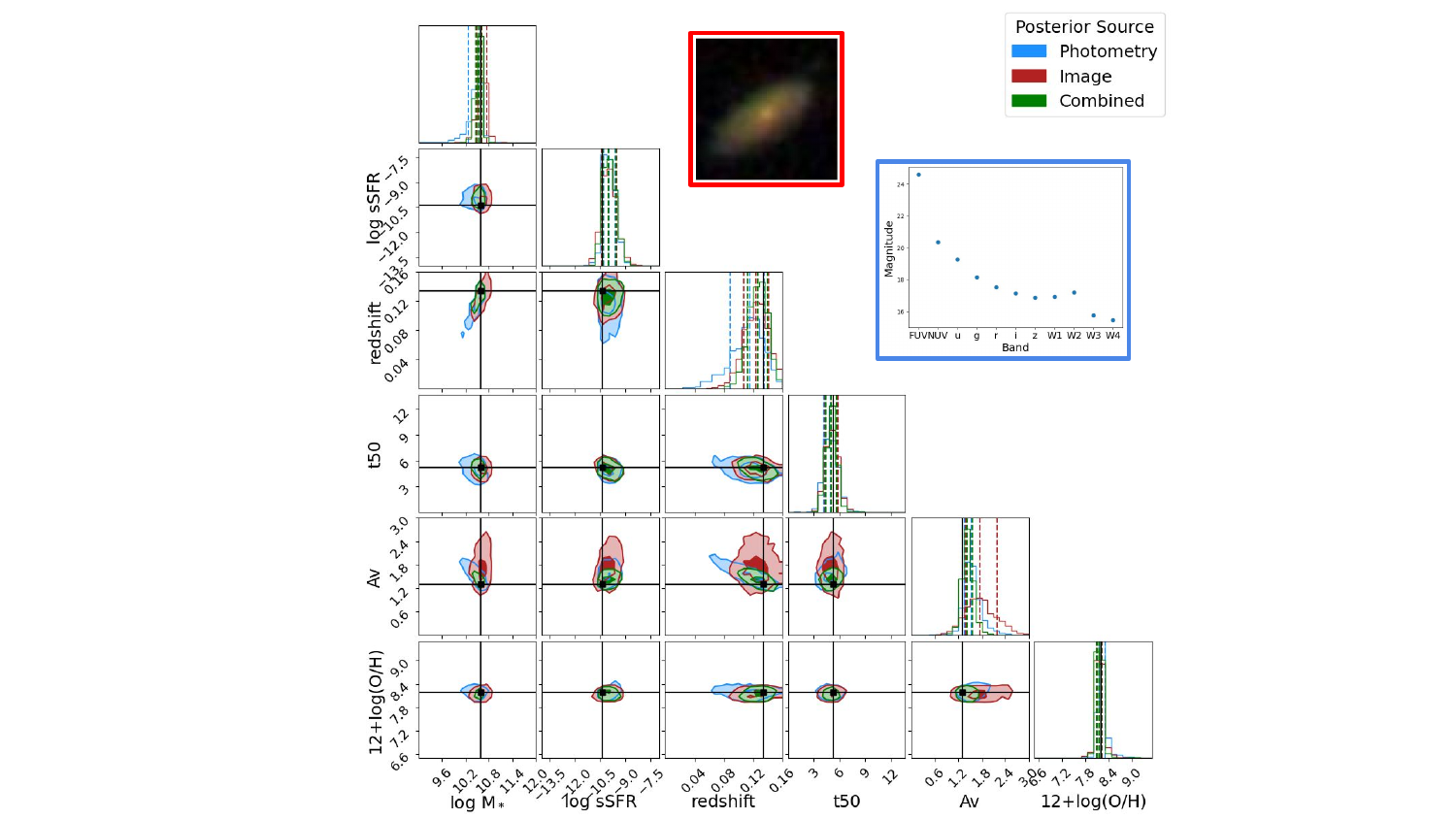}
    \caption{The posteriors generated by the image model (red), 11-band photometry model (blue), and combined model (green) for all 6 predicted physical parameters of a randomly selected galaxy from the test sample. The 11-band photometry model shows better prediction accuracy and more constrained joint probabilities. The combined model shows the best accuracy and most constrained confidence intervals, but the difference from the 11-band photometry model is marginal. This suggests that morphology does not contain any significant additional information than what is accessible from 11-band photometry.}
    \label{fig:11_band_posteriors}
\end{figure}

To compare the posteriors and joint distributions of the image model, the 11-band photometry model, and the combined model, we create a corner plot with their predictions. This can be seen in \cref{fig:11_band_posteriors}. The posteriors and joint distributions are colored depending on which model (or which input types) generated them. Red indicates predictions made using exclusively imaging data, blue indicates predictions made using exclusively 11-band photometry data, and green indicates predictions made using both imaging and 11-band photometric data.

\section{Degeneracies for the Full Population} \label{app:full_degen}

\begin{figure*}
    \centering
    \includegraphics[width=1\linewidth]{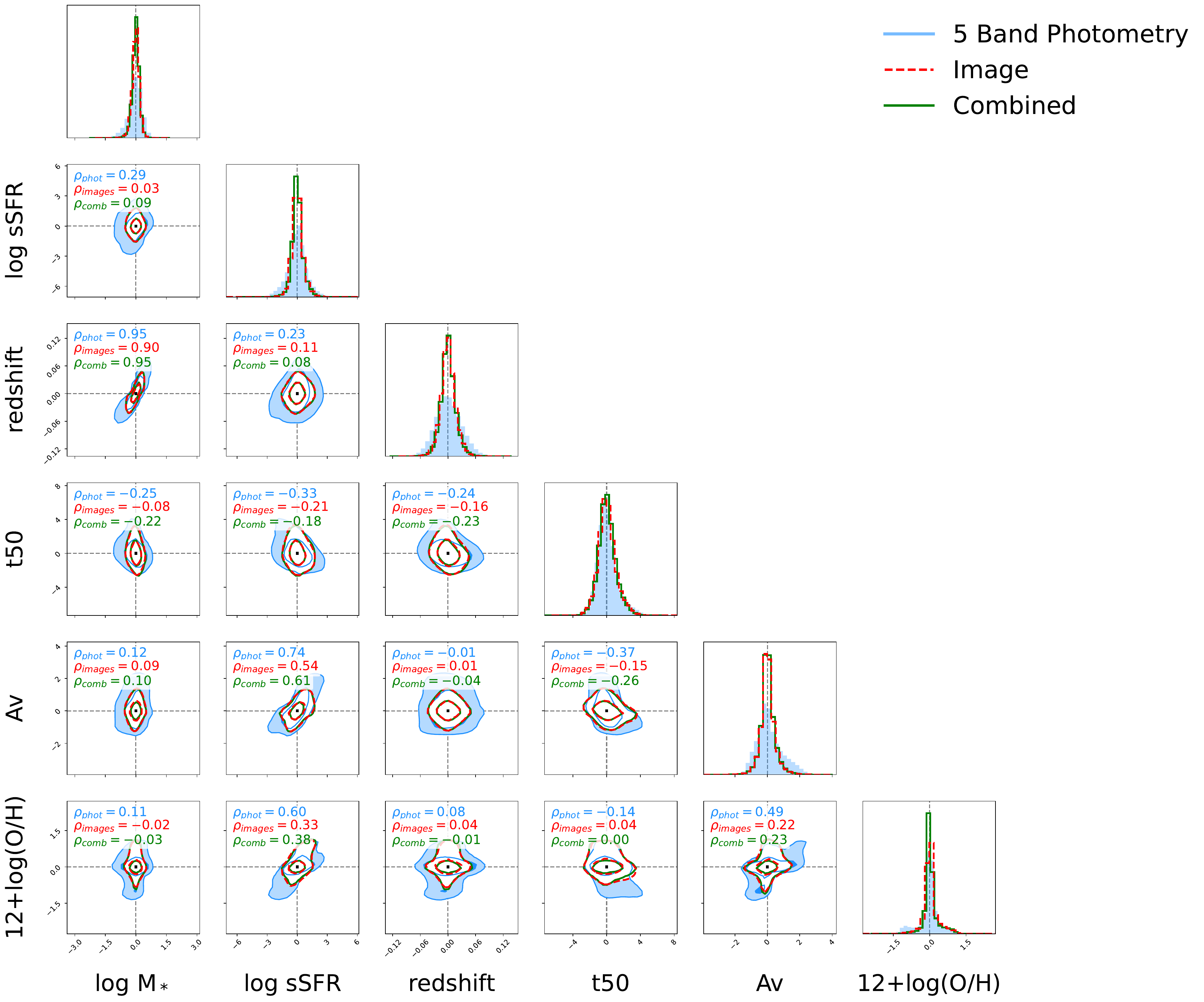}
    \caption{A population-level average inferred posteriors of the 6 physical parameters discussed in this work from the image model (red), the 5-band photometry model (blue), and combined model (green). The Spearman rank correlation coefficient ($\rho$) for each parameter pair, and for each model, is shown in the top left corner of each joint probability panel.}
    \label{fig:5_bands_covariance}
\end{figure*}

The set of three grid plots in \cref{fig:5_bands_covariance} represents the degeneracies found in the 5-band photometry model, image model, and combined model, respectively. The absolute value of $\rho$ is plotted for each parameter pair.

\section{Individual Galaxies SFHs with Confidence Intervals} \label{app:indiv_sfh}

\begin{figure*}
    \centering
    \includegraphics[width=1\linewidth]{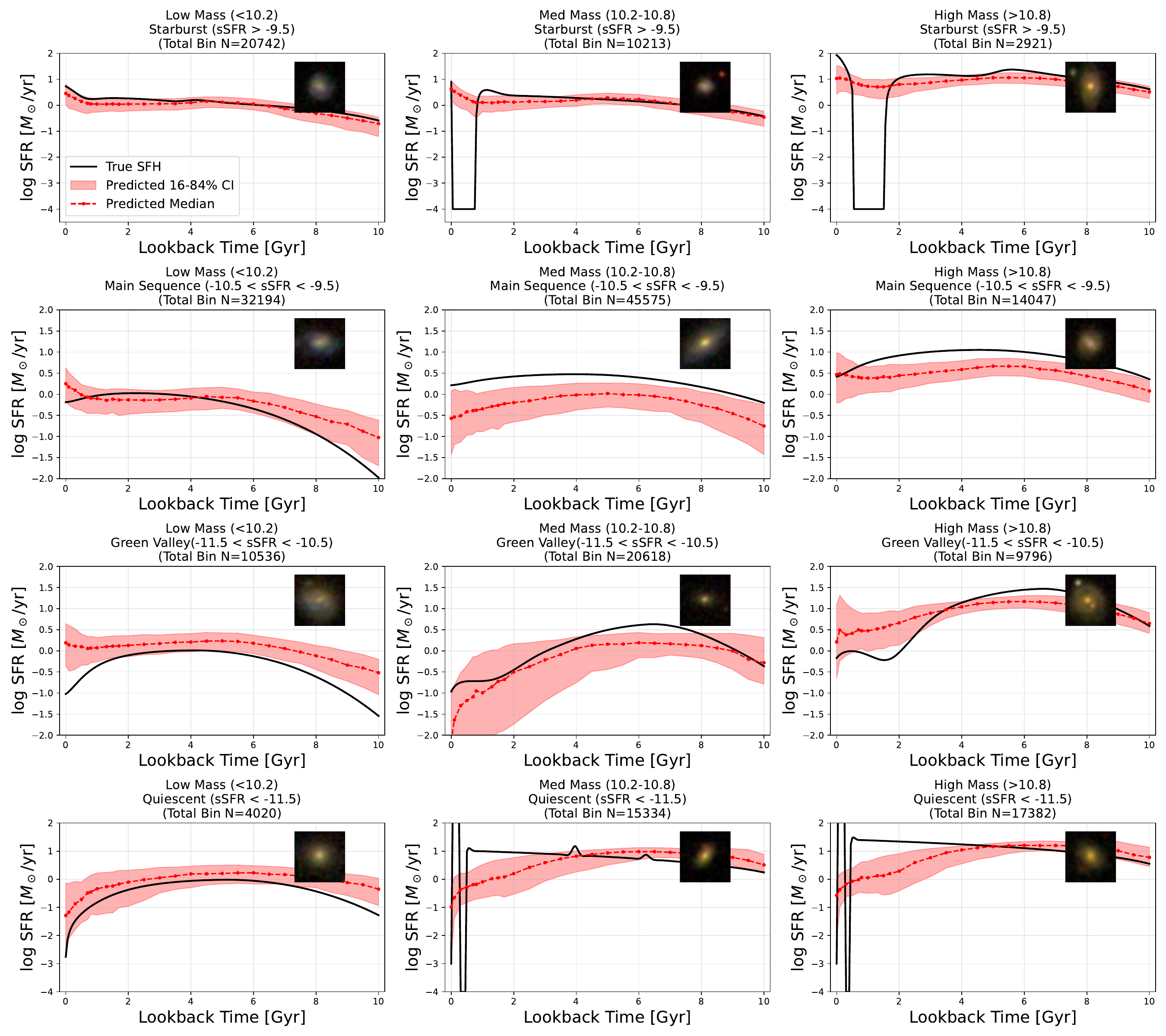}
    \caption{The accuracy of predicted SFHs and image cutouts for individual galaxies in different parts of M$_*$-SFR space. Galaxies are selected randomly from 3 mass bins and 4 sSFR bins. The true SFH history for each galaxy is plotted in black, while the predicted SFHs are plotted in red. The 16th and 84th confidence intervals for the predicted SFHs are plotted as the red shaded areas.}
    \label{fig:indv_sfh}
\end{figure*}

The accuracy of predicted SFHs and image cutouts for randomly selected individual galaxies in different parts of M$_*$-SFR space can be seen in \cref{fig:indv_sfh}. The true SFH history for each galaxy is plotted in black, while the predicted SFHs are plotted in red. The 16th and 84th confidence intervals for the predicted SFHs are plotted as the red shaded areas.

\section{Robust Memory} \label{app:robust_memory}

The maximum robust morphological memory (as outlined in \cref{sec:sfh}) of galaxies as a function of their position in the M$_*$-SFR plane.

\begin{figure}
    \centering
    \includegraphics[width=0.5\linewidth]{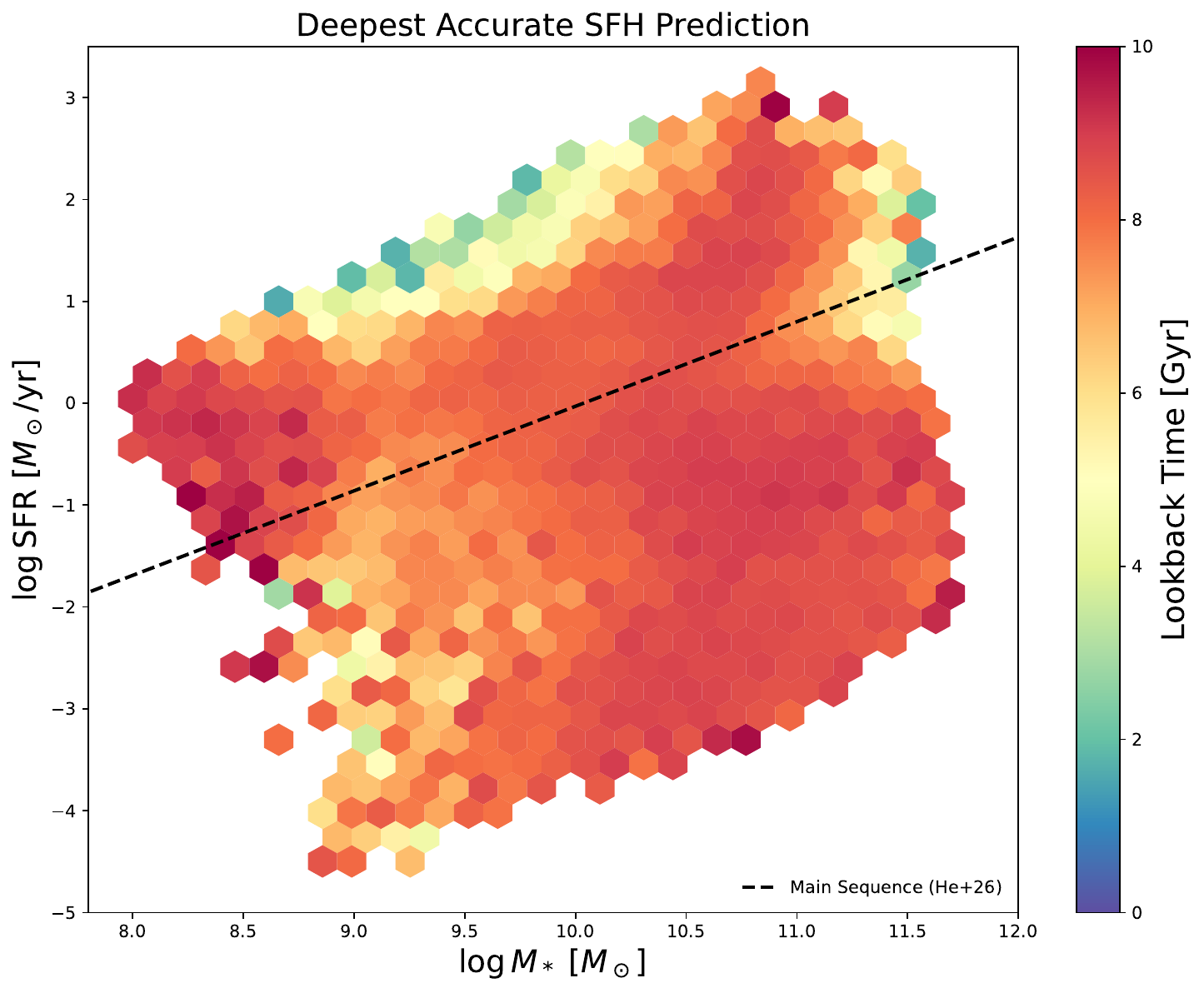}
    \caption{The maximum robust morphological memory of galaxies as a function of their position in the M$_*$-SFR plane. The hexagonal bins are colored by the deepest accurate prediction time in Gyr, representing the maximum lookback time where the predicted and actual SFHs maintain a 16–84$\%$ overlap. The dashed black line traces the star-forming main sequence from \cite{gswlc_sfms}. In contrast to their short continuous memory, starburst galaxies possess extended robust memory ($\approx$4-10 Gyr). Main sequence galaxies show long robust memories ($\approx$7-9 Gyr), complementary to their continuous memory. Green valley galaxies show much longer robust memories ($\approx$7-9 Gyr) compared to their continuous memories. Quiescent galaxies display long robust memory ($\approx$8-10 Gyr), although some quiescent galaxies show truncated robust memories.}
    \label{fig:robust_memory}
\end{figure}

\section{SFH Predicted from 5-band and 11-band photometry models} \label{app:color_sfh}

{The continuous morphological memory of galaxies mapped across the M$_*$-SFR plane coming from 11-band photometry SFH model and 5-band photometry SFH model.

\begin{figure}
    \centering
    \includegraphics[width=0.4\linewidth]{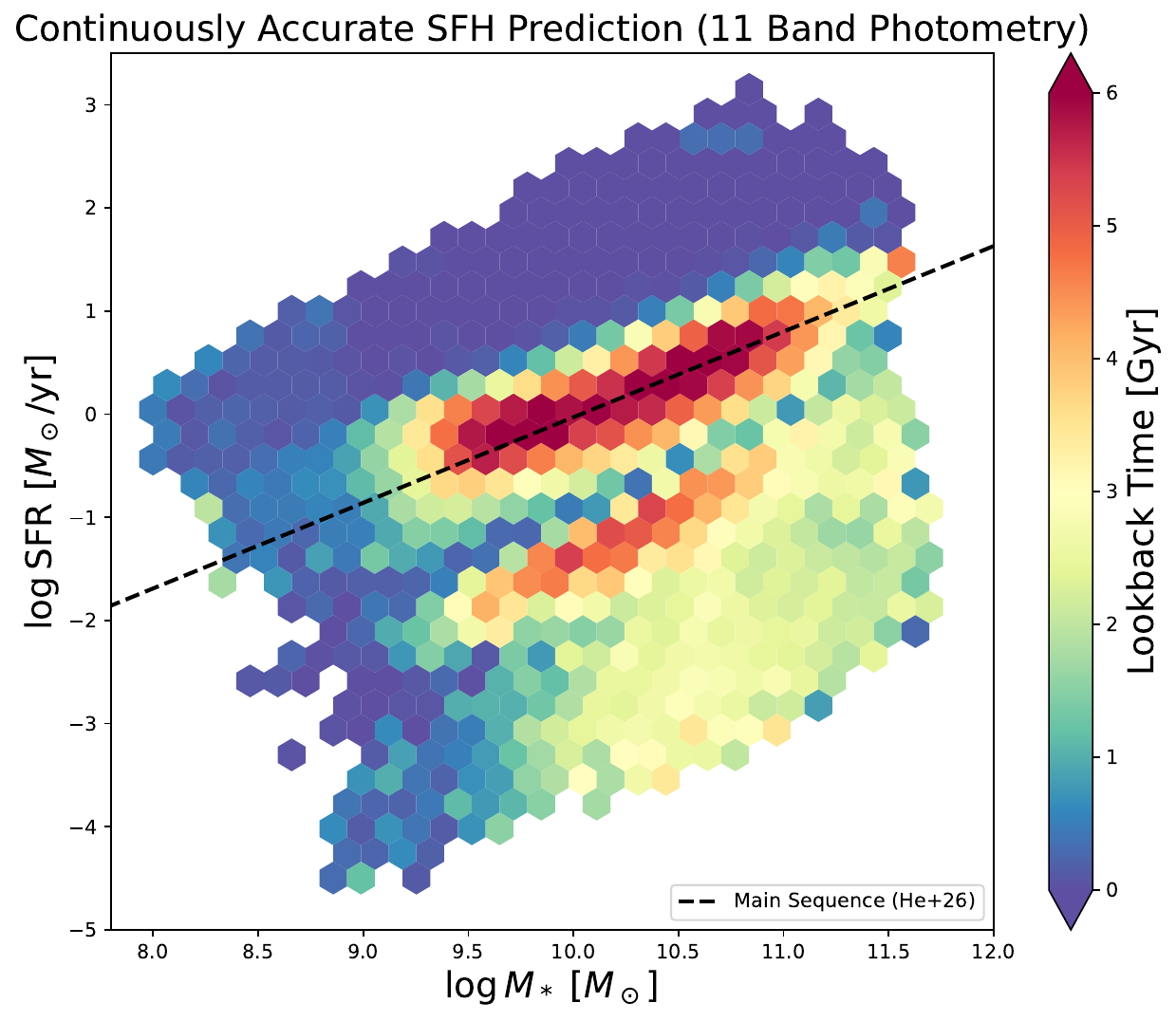}
    \includegraphics[width=0.4\linewidth]{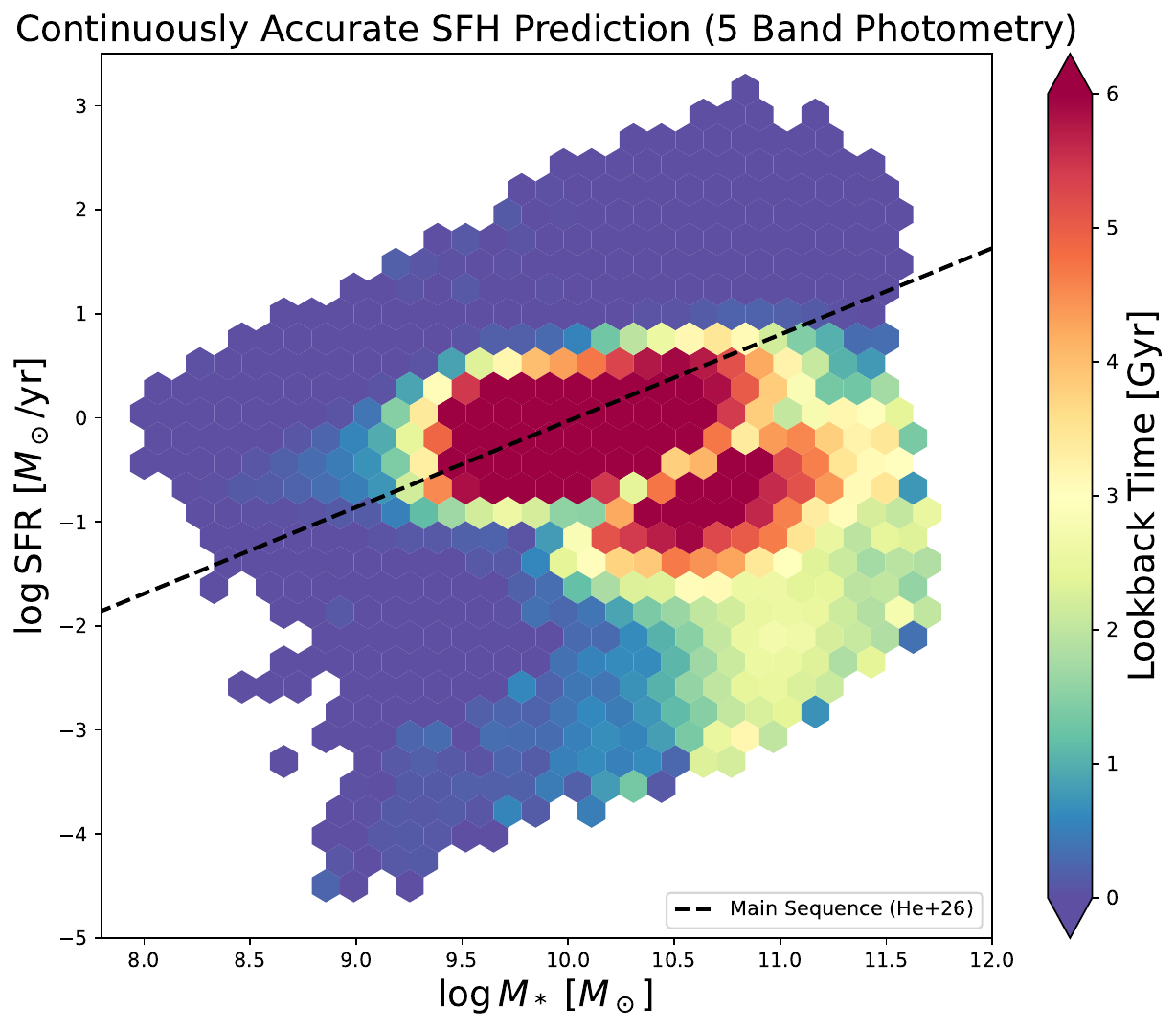}
    \caption{The continuous morphological memory of galaxies mapped across the M$_*$-SFR plane coming from 11-band photometry SFH model (left) and 5-band photometry SFH model (right). The color scale denotes the continuous lookback time in Gyr over which the structural state of a galaxy remains an accurate record of SFH, defined by a 16–84$\%$ confidence interval. The star-forming main sequence from \cite{gswlc_sfms} is indicated by the dashed black line.}
    \label{fig:cont_memory_colors}
\end{figure}

\end{document}